
\newcount\mgnf\newcount\tipi\newcount\tipoformule
\newcount\aux\newcount\aaux\newcount\pie
\newcount\greco\newcount\tiporif

\greco=0         
\mgnf=2          
\tipoformule=1   

\aux=1           

      \def\9#1{\ifnum\aux=1#1\else\relax\fi}
\aaux=0          
\pie=3           

\ifnum\mgnf=0
   \magnification=\magstep0
\hsize=14.5truecm\vsize=21.5truecm
   \parindent=4.pt\fi
\ifnum\mgnf=1
   \magnification=\magstep1
\hsize=16.0truecm\vsize=22.5truecm\baselineskip14pt\vglue6.3truecm
   \parindent=4.pt\fi
\ifnum\mgnf=2
   \magnification=\magstep2
\hsize=16.0truecm\vsize=22.5truecm\baselineskip14pt\voffset-2mm
   \parindent=0.pt\fi

\let\a=\alpha   \let\g=\gamma    \let\d=\delta \let\e=\varepsilon
  \let\h=\eta   \let\th=\vartheta\let\k=\kappa \let\l=\lambda
\let\m=\mu    \let\n=\nu    \let\x=\xi       \let\p=\pi    \let\r=\rho
\let\s=\sigma \let\t=\tau    \let\f=\varphi\let\ch=\chi
\let\ps=\psi  \let\o=\omega    
 \let\D=\Delta    
    \let\Zi=\Sigma        \let\O=\Omega

{\count255=\time\divide\count255 by 60 \xdef\oramin{\number\count255}
        \multiply\count255 by-60\advance\count255 by\time
   \xdef\oramin{\oramin:\ifnum\count255<10 0\fi\the\count255}}
\def\ora{\oramin }

\def\data{\number\day/\ifcase\month\or gennaio \or febbraio \or marzo \or
aprile \or maggio \or giugno \or luglio \or agosto \or settembre
\or ottobre \or novembre \or dicembre \fi/\number\year;\ \ora}

\setbox200\hbox{$\scriptscriptstyle \data $}

\newcount\pgn \pgn=1
\def\foglio{\number\numsec:\number\pgn
\global\advance\pgn by 1}
\def\foglioa{A\number\numsec:\number\pgn
\global\advance\pgn by 1}


\global\newcount\numsec\global\newcount\numfor
\global\newcount\numfig\global\newcount\numcap

\gdef\profonditastruttura{\dp\strutbox}
\def\senondefinito#1{\expandafter\ifx\csname#1\endcsname\relax}
\def\SIA #1,#2,#3 {\senondefinito{#1#2}%
\expandafter\xdef\csname #1#2\endcsname{#3}\else%
\write16{???? ma #1,#2 e' gia' stato definito !!!!}\fi}

\def \FU(#1)#2{\SIA fu,#1,#2 }

\def\etichetta(#1){(\veroparagrafo.\veraformula)
\SIA e,#1,(\veroparagrafo.\veraformula)
 \global\advance\numfor by 1
\9{\write15{\string\FU (#1){\equ(#1)}}}
\9{ \write16{ EQ \equ(#1) == #1  }}}

\def\etichettacap(#1){\verocapitolo
\SIA e,#1,(\verocapitolo)
\9{\write15{\string\FU (#1){\equ(#1)}}}
\9{ \write16{ EQ \equ(#1) == #1  }}}

\def\etichettapar(#1){\verocapitolo.\veroparagrafo
\SIA e,#1,{\verocapitolo.\veroparagrafo}
\9{\write15{\string\FU (#1){\parrif(#1)}}}}

\def\etichettaa(#1){(\verocapitolo.A\veroparagrafo.\veraformula)
 \SIA e,#1,(\verocapitolo.A\veroparagrafo.\veraformula)
 \global\advance\numfor by 1
\9{\write15{\string\FU (#1){\equ(#1)}}}
\9{ \write16{ EQ \equ(#1) == #1  }}}

\def\getichetta(#1){Fig. \verafigura
 \SIA e,#1,{\verafigura}
 \global\advance\numfig by 1
\9{\write15{\string\FU (#1){\equ(#1)}}}
\9{ \write16{ Fig. \equ(#1) ha simbolo  #1  }}}

\newdimen\gwidth
\def\BOZZA{
\def\alato(##1){
 {\vtop to \profonditastruttura{\baselineskip
 \profonditastruttura\vss
 \rlap{\kern-\hsize\kern-1.2truecm{$\scriptstyle##1$}}}}}
\def\galato(##1){ \gwidth=\hsize \divide\gwidth by 2
 {\vtop to \profonditastruttura{\baselineskip
 \profonditastruttura\vss
 \rlap{\kern-\gwidth\kern-1.2truecm{$\scriptstyle##1$}}}}}
}
\def\alato(#1){}
\def\galato(#1){}
\def\verocapitolo{\number\numcap}
\def\veroparagrafo{\number\numsec}\def\veraformula{\number\numfor}
\def\verafigura{\number\numfig}
\def\geq(#1){\getichetta(#1)\galato(#1)}
\def\Eq(#1){\eqno{\etichetta(#1)\alato(#1)}}
\def\eq(#1){\etichetta(#1)\alato(#1)}
\def\Eqa(#1){\eqno{\etichettaa(#1)\alato(#1)}}
\def\eqa(#1){\etichettaa(#1)\alato(#1)}
\def\eqv(#1){\senondefinito{fu#1}$\clubsuit$#1\write16{Manca #1 !}%
\else\csname fu#1\endcsname\fi}
\def\equ(#1){\senondefinito{e#1}\eqv(#1)\else\csname e#1\endcsname\fi}
\def\Cap(#1){\etichettacap(#1)}
\def\Par(#1){\etichettapar(#1)}
\def\parrif(#1){\senondefinito{e#1}\eqv(#1)\else\csname e#1\endcsname\fi}

\openin13=#1.aux \ifeof13 \relax \else
\input #1.aux \closein13\fi
\openin14=\jobname.aux \ifeof14 \relax \else
\input \jobname.aux \closein14 \fi
\9{\openout15=\jobname.aux}


\newskip\ttglue%
%
%
%
%
\font\ottorm=cmr8\font\ottoi=cmmi8\font\ottosy=cmsy8%
\font\ottobf=cmbx8\font\ottott=cmtt8%
\font\ottosl=cmsl8\font\ottoit=cmti8%
\font\sixrm=cmr6\font\sixbf=cmbx6\font\sixi=cmmi6\font\sixsy=cmsy6%
\font\fiverm=cmr5\font\fivesy=cmsy5\font\fivei=cmmi5\font\fivebf=cmbx5%
\def\ottopunti{\def\rm{\fam0\ottorm}%
\textfont0=\ottorm\scriptfont0=\sixrm\scriptscriptfont0=\fiverm%
\textfont1=\ottoi\scriptfont1=\sixi\scriptscriptfont1=\fivei%
\textfont2=\ottosy\scriptfont2=\sixsy\scriptscriptfont2=\fivesy%
\textfont3=\tenex\scriptfont3=\tenex\scriptscriptfont3=\tenex%
\textfont\itfam=\ottoit\def\it{\fam\itfam\ottoit}%
\textfont\slfam=\ottosl\def\sl{\fam\slfam\ottosl}%
\textfont\ttfam=\ottott\def\tt{\fam\ttfam\ottott}%
\textfont\bffam=\ottobf\scriptfont\bffam=\sixbf%
\scriptscriptfont\bffam=\fivebf\def\bf{\fam\bffam\ottobf}%
\tt\ttglue=.5em plus.25em minus.15em%
\setbox\strutbox=\hbox{\vrule height7pt depth2pt width0pt}%
\normalbaselineskip=9pt\let\sc=\sixrm\normalbaselines\rm}
\catcode`@=11
\def\footnote#1{\edef\@sf{\spacefactor\the\spacefactor}#1\@sf%
\insert\footins\bgroup\ottopunti%
\interlinepenalty100\let\par=\endgraf%
\leftskip=0pt\rightskip=0pt%
\splittopskip=10pt plus 1pt minus 1pt\floatingpenalty=20000%
\smallskip\item{#1}\bgroup\strut\aftergroup\@foot\let\next}%
\skip\footins=12pt plus 2pt minus 4pt%
\dimen\footins=30pc%
\catcode`@=12%
\let\nota=\ottopunti%

%
%
%
\newdimen\xshift \newdimen\xwidth \newdimen\yshift \newdimen\ywidth

\def\ins#1#2#3{\vbox to0pt{\kern-#2\hbox{\kern#1 #3}\vss}\nointerlineskip}

\def\eqfig#1#2#3#4#5{
\par\xwidth=#1 \xshift=\hsize \advance\xshift
by-\xwidth \divide\xshift by 2
\yshift=#2 \divide\yshift by 2
\line{\hglue\xshift \vbox to #2{\vfil
#3 \includegraphics{#4.ps}
}\hfill\raise\yshift\hbox{#5}}}

\def\eqfigbis#1#2#3#4#5#6#7{
\par\xwidth=#1 \multiply\xwidth by 2
\xshift=\hsize \advance\xshift
by-\xwidth \divide\xshift by 3
\yshift=#2 \divide\yshift by 2
\ywidth=#2
\line{\hglue\xshift
\vbox to \ywidth{\vfil #3 \includegraphics{#4.ps}}
\hglue30pt
\vbox to \ywidth{\vfil #5 \includegraphics{#6.ps}}
\hfill\raise\yshift\hbox{#7}}}

\def\dimenfor#1#2{\par\xwidth=#1 \multiply\xwidth by 2
\xshift=\hsize \advance\xshift
by-\xwidth \divide\xshift by 3
\divide\xwidth by 2
\yshift=#2 
\ywidth=#2}


\def\eqfigter#1#2#3#4#5#6#7{
\line{\hglue\xshift
\vbox to \ywidth{\vfil #1 \includegraphics{#2.ps}}
\hglue30pt
\vbox to \ywidth{\vfil #3 \includegraphics{#4.ps}}\hfill}
\multiply\xshift by 3 \advance\xshift by \xwidth \divide\xshift by 2
\line{\hfill\hbox{#7}}
\line{\hglue\xshift
\vbox to \ywidth{\vfil #5 \includegraphics{#6.ps}}\hfill}}


\def\8{\write12}  


\def\V#1{{\bf#1}}
\def\T#1{#1\kern-4pt\lower9pt\hbox{$\widetilde{}$}\kern4pt{}}
\let\dpr=\partial\def\Dpr{{\V\dpr}}
\let\io=\infty\let\ig=\int
\def\fra#1#2{{#1\over#2}}\def\media#1{\langle{#1}\rangle}\let\0=\noindent

\def\guida{\leaders\hbox to 1em{\hss.\hss}\hfill}
\def\tende#1{\,\vtop{\ialign{##\crcr\rightarrowfill\crcr
              \noalign{\kern-1pt\nointerlineskip}
              \hglue3.pt${\scriptstyle #1}$\hglue3.pt\crcr}}\,}
\def\otto{\ {\kern-1.truept\leftarrow\kern-5.truept\to\kern-1.truept}\ }

\def\pagina{\vfill\eject}
\def\txt{\textstyle}
\def\st{\scriptscriptstyle}
\def\*{\vskip0.2truecm}

\def\eg{\hbox{\it e.g.\ }}
\def\ap{\hbox{\it a priori\ }}
\def\ie{\hbox{\it i.e.\ }}\def\cfr{{\it c.f.r.\ }}
\def\fiat{{}}

\def\\{\hfill\break}
\def\={{\,\equiv\,}}
\def\annota#1{\footnote{$\bf{}^#1$}}\let\ch=\chi
\def\defi{\,{\buildrel def \over =}\,}

\ifnum\aux=1\BOZZA\else
\let\Eq=\eqno\def\eq{}\let\Eqa=\eqno\def\eqa{}\def\Par(S#1){#1}
\def\parrif(S#1){#1}
\def\equ{{}}\footline={\hss\tenrm\folio\hss}
\relax\fi
\ifnum\tipoformule=0\let\Eq=\eqno\def\eq{}\let\Eqa=\eqno\def\eqa{}
\def\equ{{}}\footline={\hss\tenrm\folio\hss}\fi
\ifnum\aaux=0\let\foglio=\folio\else\relax\fi

\def\BB{{\cal B}}

\def\FF{{\cal F}}

\def\\{\hfill\break}
\let\dpr=\partial\let\io=\infty
\let\ig=\int
\def\media#1{\langle{#1}\rangle}

\def\guida{\leaders\hbox to 1em{\hss.\hss}\hfill}

\def\pagina{\vfill\eject}

\def\eg{\hbox{\it e.g.\ }}
\def\ap{\hbox{\it a priori\ }}

\def\fiat{{}}
\def\Dpr{{\V\dpr}}
\def\={{\,\equiv\,}}
\let\ch=\chi  

\def\st{\scriptscriptstyle}
\def\fra#1#2{{#1\over#2}}\def\*{\vskip0.3truecm}

\let\th=\vartheta

\def\pagina{\vfill\eject}

\def\V#1{{\underline#1}}
\def\xx{{\V x}}\def\yy{{\V y}}\def\kk{{\V k}}
\def\yy{{\V y}}

\def\2{{1\over2}}

\def\T#1{{#1_{\kern-3pt\lower7pt\hbox{$\widetilde{}$}}\kern3pt}}
\def\VV#1{{\underline #1}_{\kern-3pt
\lower7pt\hbox{$\widetilde{}$}}\kern3pt\,}

\newbox\strutboxa
\setbox\strutboxa=\hbox{\vrule height8.5pt depth2.25pt width0pt}
\def\struta{\relax\ifmmode\copy\strutboxa\else\unhcopy\strutboxa\fi}
\def\W#1{#1_{\kern-3pt\lower7.5pt\hbox{$\widetilde{}$}}\kern2pt\,\struta}

\def\CC{{\cal C}}
\def\FF{{\cal F}}

\def\xx{{\V x}}

\def\={\equiv}

\let\txt=\textstyle

\def\lis{\,\overline}



\def\ifnextchar#1#2#3{\let\tempe #1\def\tempa{#2}\def\tempb{#3}\futurelet
\tempc\ifnch}
\def\ifnch{\ifx\tempc\tempe\let\tempd\tempa\else\let\tempd\tempb\fi\tempd}
\def\gobble#1{}
\ifnum\greco=1\font\tengr=grreg10\else\font\tengr=cmmi10\fi
\def\greekmode{%
\catcode`\<=13
\catcode`\>=13
\catcode`\'=11
\catcode`\`=11
\catcode`\~=11
\catcode`\"=11
\catcode`\|=11
\lccode`\<=`\<%
\lccode`\>=`\>%
\lccode`\'=`\'%
\lccode`\`=`\`%
\lccode`\~=`\~%
\lccode`\"=`\"%
\lccode`\|=`\|%
\tengr\def\bf{\tengrbf}\def\tt{\tengrtt}}
\newcount\vwl
\newcount\acct
\def\lt{<}

{
  \greekmode
  \gdef>{\ifnextchar `{\expandafter\smoothgrave\gobble}{\char\lq\>}}
  \gdef<{\ifnextchar `{\expandafter\roughgrave\gobble}{\char\lq\<}}
  \gdef\smoothgrave#1{\acct=\rq137 \vwl=\lq#1 \dobreathinggrave}
  \gdef\roughgrave#1{\acct=\rq103 \vwl=\lq#1 \dobreathinggrave}
  \gdef\dobreathinggrave{\ifnum\vwl\lt\rq140	
    \char\the\acct\char\the\vwl\else\expandafter\testiota\fi}
      \gdef\testiota{\ifnextchar |{\addiota\doaccent\gobble}{\doaccent}}
        \gdef\addiota{\ifnum\vwl=\lq a\vwl=\rq370
            \else\ifnum\vwl=\lq h\vwl=\rq371 \else\vwl=\rq372 \fi\fi}
              \gdef\doaccent{\accent\the\acct \char\the\vwl\relax}
              }

\newif\ifgreek\greekfalse

\def\begingreek{\bgroup\greektrue\greekmode}
\def\endgreek{\egroup}

\let\math=$
{\catcode`\$=13
}

\relax
\begingreek

\endgreek
\relax

\ifnum\greco=0

\relax
\fi

\headline{\ifodd\pageno\hss\tit\hss\number\pageno
\else
\number\pageno\hss\tit\hss\fi}

\def\tit{{}}
\footline={\rlap{\hbox{\copy200}}\hss}

\relax

\tolerance=9000

%
\def\rigorde#1#2{\rlap{\kern#1\raise#2\hbox{\vbox{\hrule width 5pt}}}}
\def\rigorsi#1#2{\rlap{\kern-5pt\kern#1\raise#2\hbox{\vbox{\hrule width 5pt}}}}
\def\rigveal#1#2{\rlap{\kern#1\raise#2\hbox{\vrule height5pt depth 0pt}}}
\def\rigveba#1#2{\rlap{\kern#1\raise#2\hbox{\vrule height0pt depth 5pt}}}
\newdimen\sopra \newdimen\sotto \newdimen \destra \newdimen\sinistra
\newdimen\shiftvert\newdimen\shiftor\newdimen\alth
\shiftvert=9.25in \advance\shiftvert by -\vsize \divide\shiftvert by 2
\shiftor=6.125in \advance\shiftor by -\hsize \divide\shiftor by 2
\sopra=-\baselineskip \advance\sopra by -\baselineskip \advance\sopra by 10pt
\sotto=\sopra
\advance\sopra by \shiftvert
\alth=\sopra \advance\alth by 5pt
\advance\sotto by -\vsize \advance\sotto by -\shiftvert
\destra=\hsize \advance\destra by \shiftor
\sinistra=0pt \advance\sinistra by -\shiftor
\def\cornice{%
\rigorsi{\sinistra}{\sopra}\rigveal{\sinistra}{\sopra}%
\rigorde{\destra}{\sopra}\rigveal{\destra}{\sopra}%
\rigorsi{\sinistra}{\sotto}\rigveba{\sinistra}{\sotto}%
\rigorde{\destra}{\sotto}\rigveba{\destra}{\sotto}%
 }
%
%

\def\makeheadline{\vbox to 0pt{\vskip10pt\vskip-\baselineskip
\vskip-\baselineskip\vskip-\alth
\line{\vbox to8.5pt{}\cornice
\the\headline}\vss}\nointerlineskip}


\ifnum\pie=0\footline={\rlap{\hbox{\copy200}\hss\tenrm
\folio\hss}}\fi
\ifnum\pie=1\footline={\hss\tenrm\folio\hss}\fi
\ifnum\pie=2\footline={\rlap{\hbox{\copy200}\ $\st[\number\pageno]$}\hss\tenrm
\foglio\hss}\fi
\ifnum\pie=3\nopagenumbers\fi


%
%
%

\footline={\rlap{\hbox{\copy200}}\hss}

\newcount\numsec\newcount\numfor\newcount\numapp

\def\1{\ifnum\mgnf=0\vfill\eject\else\relax\fi}
\def\Idea{{\it Idea:\ }}\def\mbe{{\\*\hfill\hbox{\it
mbe\kern0.5truecm}}\vskip3.truept}
\def\indica{\leaders \hbox to 0.5cm{\hss.\hss}\hfill}

\def\2{}\def\3{\ifnum\mgnf=0\pagina\hbox{}\fi}
\def\4{}

\def\4{\ifnum\mgnf=0\pagina\hbox{}\fi}
\def\2{}\def\3{}

\def\fiat{{}}
\language=0


\fiat

\def\tit{\nota CKN theory of singularities}
\numcap=1

\vglue1.truecm
\centerline{\bf CKN theory of singularities  of} 
\centerline{\bf weak solutions of the Navier-Stokes equations}
\centerline{\it Giovanni Gallavotti}
\centerline{Fisica, Roma1 and I.N.F.N.}
\*\*

\0{\bf Abstract: A series of five lectures delivered at the
C.I.M.E. course on ``Mathematical foundations of turbulent viscous
flows'', 1-6 september, 2003}
\*\*

\0{\bf \S0. Notations.}
\numsec=0\numfor=1\* 
\*

The lectures are devoted to a complete exposition of the theory of
singularities of the Navier Stokes equations solution studied by
Leray, in a simple geometrical setting in which the fluid is enclosed
in a container $\O$  with periodic boundary conditions and side size $L$.
The theory is due to the work of Scheffer, Caffarelli, Kohn, Nirenberg
and is called here CKN-theory as it is inspired by the work of the last
three authors which considerably improved the earlier estimates of
Scheffer.

Although the theory of Leray is well known I recall it here getting at
the same time a chance at establishing a few notations, [Ga02].

\0(1) An underlined letter , \eg $\V A, \W A,\ldots$, denotes a
$3$--dimensional vector (\ie three real or complex numbers) and
underlined partial derivative symbol $\V \dpr, \W\dpr,\ldots$ denotes
the gradient operator $(\dpr_1,\dpr_2,\dpr_3)$. A vector field $\V u$
is a function on $\O$.

\0(2) repeated labels convention is used (labels are letters or other)
when not ambiguous: hence $ \V A\cdot \V B$ or $\W A\cdot\W B$ means
sum over $i$ of $A_iB_i$. Therefore $\V\dpr\cdot\V u$, if $\V u$ is a
vector field, is the divergence of $\V u$, namely $\sum_i\dpr_i
u_i$. $\V\dpr\cdot\V\dpr=\D$ is the Laplace operator.

\0(3) Multiple derivatives are tensors, so that $\V\dpr\W\dpr f$ is
the tensor $\dpr_{ij}f$. The $L_2(\O)$ is the space of the square
integrable functions on $\O$: the squared norm of $f\in L_2$ will be
$||f||_2^2\defi \ig_\O |f(\V x)|^2s\V x$.

\0(4) The Navier-Stokes equation with regularization parameter $\l$ is

$$\dot\V u=\n\D\V u-\media{\W u}_\l\cdot\W\dpr\,\V u- \V\dpr p,
\qquad\Dpr\cdot\V u=0,\qquad\ig_\O\V u\,d\V x=\V0\Eq(1.1)$$
where the unkowns are $\V u(\V x,t),p(\V x,t)$ with zero average and, 
[Ga02],

(i) $\V u$ is a divergenceless field, $p$ is a scalar field
\\
(ii) $\media{\V u}_\l=\ig_\O \ch_\l(\V x-\V y) \V u(\V y) d\V y$ and
$\ch_\l$ is defined in terms of a $C^\io(\O)$ function
$\xx\to\chi(\xx)\ge0$ not vanishing in a small
neighborhood of the origin and with integral $\ig\chi(\xx)\,
d\xx\=1$: the function $\ch(\xx)$ can be regarded as a periodic
function on $\O$ or as a function on $R^3$ with value $0$ outside
$\O$, as we shall imagine that $\O$ is centered at the origin, to fix
the ideas. For $\l\ge1$ also the function
$\chi_\l(\xx)\defi\l^3\chi(\l\xx)$ can be regarded as a periodic
function on $\O$ or as a function on $R^3$: it is an ``approximate
Dirac's $\d$--function''. Usually the NS equation contains a volume 
force too: here we set it equal to $\V 0$.\annota{1}{\nota This is a
simplicity assumption as the extension of the theory to cases with
time independent smooth (\eg $C^\io$) volume forces would be immediate
and just a notational nuisance.}
\\
(iii) The initial datum is a divergenceless velocity field $\V u^0\in
L_2(\O)$ with $\V 0$ average; no
initial datum for $p$ as $p$ is determined from $\V u^0$.

\0(5) A {\it weak solution} of the NS equations with initial datum $\V
u_0\in L_2(\O)$ is a limit on subsequences of $\l\to\io$ of solutions
$\V u^\l, p^\l$ of \equ(1.1). This means that the Fourier transforms
of $\V u^\l$, and $p^\l$ exist and have components $\V u^\l_{\V
k}(t)\=\ig_\O e^{-i\V k\dot\V x} \V u^\l(\V x) d\V x$ ($\V
k=\fra{2\p}L\V n$, $\V n\in Z^3$), and $p^\l_\V k(t)$ which have a limit
as $\l\to\io$ (on subsequences) for each $\V k\ne\V 0$ and the limit
of the $\V u_{\V k}(t)$ is absolutely continuous. This is equivalent
to the existence of the limits of the $L_2$ products $(\V u(t), \V
\f)_{L_2}$ and $(\V p(t), \psi)_{L_2}$ for all $t\in (0,\io]$ and for
all test functions $\V f(\V x),\ps(\V x)$.  There might be several
such limits (\ie the limit may depend on the subsequence) and what
follows applies to any one among them.  
\*

The core of the analysis will deal with the regularized equation and
the properties of its solutions, which are easily shown to be $C^\io$
in $\V x\in \O$ and in $t\in(0,\io)$ if the intial datum is $\V u\in
L_2(\O)$. The limit $\l\to\io$ will be taken at the end and it is
where the theory becomes non conctructive because there is need to
consider the limit on subsequences. 

Of course the point is to obtain bounds which are uniform in
$\l\to\io$ and the limit $\l\to\io$ only intervenes at the end to
formulate the results in a nice form.

The theory of Leray is based on the following \ap bounds, see
section 3.2 in [Ga02], on solutions of \equ(1.1)  with
initial datum $\V u_0$ with $L_2$ square norm $E_0$
$$||\V u^{\l}(t)||^2_2\le E_0,\qquad
\ig^t_0\,d\t||\V\dpr {\W u}^{\l}(\t)||_2^2\le
\fra12\,{E_0\,\n^{-1}}\Eq(1.2)$$
satisfied by the solution $\V u^{\l}$.

The notes are extracted from reference [Ga02] to which the reader is
referred for details on the above results and have been made
independent from [Ga02] modulo the above results (in fact, essentially, only
modulo the statements on the regularized equation \equ(1.1)).

The proof is conceptually quite simple and is based on a few (clever)
\ap Sobolev inequalities: the estimates are discussed in Sect. 1-3,
which form an introduction.

Their application to the analysis of \equ(1.1) is in Sect. 4 where the
main theorems are discussed and the CKN main result is reduced via the
inequalities to Scheffer's theorem. The method is a kind of multiscale
analysis which allows us to obtain regularity {\it provided} a control
quantity, identified here as the ``{\it local Reynolds number}'' on
various scales, is small enough. Unfortunately it is not (yet)
possible to prove that the local Reynolds number is small on all small
regions (physically this would mean that in such regions the flow
would be laminar, hence smooth on small scale). 

{\it However} the \ap bounds give the information that the local
Reynolds number must be small near many points in $\O$ and, via
standard techniques, an estimate of the dimension of the possibly bad
points follows.  The application to the fractal dimension bound is
essentially an ``abstract reasoning'' consequence of the results of
Sect. 4 and is in Sect. 5. 

The proofs of the various Sobolev inequalities necessary to obtain the
key Scheffer's theorem and of the new ones studied by CKN is decribed
concisely but in full detail in the series of problems at the end of
the text: the hints describe quickly the various steps of the proofs
(however without skipping any detail, to my knowledge).  
\*

\0{\bf \S1. Leray's solutions and energy.}
\numsec=1\numfor=1\* 

The theory of space--time singularities will be partly based upon
simple general {\it kinematic inequalities}, which therefore have
little to do with the Navier--Stokes equation, and partly they will be
based on the local energy conservation which follows as a consequence
of the Navier--Stokes equations {\it but it is not equivalent to
them}.

Energy conservation for the regularized equations \equ(1.1) says that
the kinetic energy variation in a given volume element $\D$ of the
fluid, in a time interval $[t_0,t_1]$, plus the energy dissipated
therein by friction, equals the sum of the kinetic energy that in the
time interval $t\in [t_0,t_1]$ enters in the volume element plus the
work performed by the pressure forces (on the boundary element) plus
the work of the volume forces (none in our case). The analytic form of
this relation is simply obtained by multiplying both sides of the
first of the \equ(1.1) by $\V u$ and integrating on the volume element
$\D$ and over the time interval $[t_0,t_1]$.

The relation that one gets can be generalized to the case in which the
volume element has a time dependent shape. And an even more general
relation can be obtained by multiplying both sides of \equ(1.1) by
$\f(\xx,t)\V u(\xx,t)$ where $\f$ is a $C^\io(\O\times(0,s])$ function
with $\f(\xx,t)$ zero for $t$ near $0$ (here $s$ is a positive
parameter).

Energy conservation in a sharply defined volume $\D$ and time interval
$t\in [t_0,t_1]$ can be obtained as limiting case of choices of $\f$
in the limit in which it becomes the characteristic function of the
space--time volume element $\D\times[t_0,t_1]$.

Making use of a regular function $\f(\xx,t)$ is useful, particularly
in the rather ``desperate'' situation in which we are when using the
theory of Leray. The ``solutions'' $\V u$ (obtained by removing, in
\equ(1.1), the regularization, \ie letting $\l\to\io$) are only ``weak
solutions''. Therefore, the relations that are obtained in the limit
$\l\o\io$ can be interpreted as valid only after suitable integrations
by parts that allow us to avoid introducing derivatives of $\V u$
(whose existence is not guaranteed by the theory) at the ``expense''
of differentiating the ``test function'' $\f$.

Performing analytically the computation of the energy balance,
described above in words, in the case of the regularized equation
\equ(1.1) and via a few integrations by parts\annota{2}{\nota
The solutions of \equ(1.1) are $C^\io(\O\times[0,\io))$ so that there is
no need to justify integrating by parts.} we get the following
relation
$$\eqalign{
&{1\over2}\ig_\O d\V\x|\V u(\V\x,s)|^2\f(\V \x,s)+\n\ig_0^s d t
\ig_\O\f(\V\x,t)|\V\dpr\W u(\V\x,t)|^2d\V x =\cr
&=\ig_0^s\ig_\O\Big[{1\over2}(\f_t+\n\,\D\f)|\V u|^2+
\fra12|\V u|^2\langle\W u\rangle_\l\cdot\W\dpr\f
+p\,\V u\cdot\V\dpr\f\Big]\,dt\,d\V\x\cr}\Eq(1.3)$$
where $\f_t\=\dpr_t \f$ and $\V u=\V u^\l$ is in fact depending also
on the regularization parameter $\l$; here $p$ is the pressure
$p=-\sum_{ij} \D^{-1}\dpr_i\dpr_j (u_i u_j)$.

Suppose that the solution of \equ(1.1) with fixed initial datum $\V
u_0$ converges (weakly in $L_2$), for $\l\to\io$, to a ``Leray
solution'' $\V u$ possibly only over a subsequence $\l_n\to\io$.

The \equ(1.3) implies that any (in case of non uniqueness) Leray
solution $\V u$ verifies the
{\it energy inequality}:
$$\eqalign{
&{1\over2}\ig_\O |\V u(\V \x,s)|^2\f(\V \x,s)\,d\V \x+\n\ig_{t\le
s}\ig_\O \f(\V\x,t)|\V\dpr\W u(\V\x,t)|^2d\V \x d t\le\cr
&\le\ig_{t\le s}\ig_\O \Big[{1\over2}(\f_t+\n\,\D\f)|\V u|^2+
\fra12|\V u|^2\W u\cdot\W\dpr\f+p\,\W u\cdot\W\dpr\f\Big]
\,d\V \x\, dt\cr}\Eq(1.4)$$
where the pressure $p$ is given by $p=-\sum_{ij}\D^{-1}\dpr_i\dpr_j
(u_i u_j)\=-\D^{-1}\W\dpr\,\V\dpr (\V u\,\W u)$ $\=-\D^{-1}(\W\dpr\,\V
u)^2$.
\*

\0{\bf Remarks:} {(1) } It is important to remark that in this
relation one might expect the {\it equality sign}: as we shall see the fact
that we cannot do better than just obtaining an inequality means that
the limit necessary to reach a Leray solution can introduce a
``spurious dissipation'' that we are simply unable to understand on
the basis of what we know (today) about the Leray solutions.

{(2) } The above ``strange'' phenomenon reflects our inability to
develop a complete theory of the Navier--Stokes equation, but one can
conjecture that no other dissipation can take place and that a (yet to
come) complete theory of the equations could show this. Hence we
should take the inequality sign in \equ(1.4) as one more manifestation
of the inadequacy of the Leray's solution.
\*

The proof of \equ(1.4) and of the other inequalities that we shall
quote and use in this section is elementary and based, \cfr problem
[15] below, on a few general ``kinematic inequalities'' that we now
list (all of them will be used in the following but only (S) and (CZ)
are needed to check \equ(1.4)).
\*

\0{\bf \S2. Kinematic inequalities.}
\numsec=2\numfor=1\*

A first ``kinematic'' inequality, \ie the first inequality that we
shall need and that holds for any function $f$, is\annota{3}{\nota The
inequalities should be regarded as inequalities for $C^\io$ functions;
they can be extended to the appropriate Sobolev spaces by continuity.}
\*

\0(P) {\it Poincar\'e inequality:}
$$\ig_{B_r} d\V x\,|f-F|^\a\le C^P_\a\,r^{3-2\a}\left(\ig_{B_r}d\V x\,
|\V\dpr f|\right)^\a,\qquad1\le\a\le\fra32\Eq(2.1)$$
where $F$ is the average of $f$ on the ball $B_r$ with radius $r$ and
$C^P_\a$ is a suitable constant. We shall denote \equ(2.1) by (P).
\*

A second kinematic  inequality that we shall use is
\*

\0(S) {\it Sobolev inequality:}
$$\eqalign{
&\ig_{B_r}|\V u|^q\,d\xx \le
C^S_q\Big[\left(\ig_{B_r}(\V\dpr\W u)^2\,d\xx \right)^a\cdot\left(\ig_{B_r}|\V
u|^2\,d\xx \right)^{{q/2}-a}+\cr
&+r^{-2a}\left(\ig_{B_r}|\V u|^2\,d\xx\right)^{q/2}
\Big]\qquad {\rm if}\ 2\le q\le6,\quad a={3\over 4}(q-2)\cr}\Eq(2.2)$$
where $B_r$ is a ball of radius $r$ and the integrals are performed
with respect to $d\xx$. The $C^S_q$ is a suitable constant; the second
term of the right hand side can be omitted if $\V u$ has zero average
over $B_r$. We shall denote
\equ(2.2) by (S), [So63].
\*
A third necessary kinematic inequality will be
\*
\0(CZ) {\it Calderon--Zygmund inequality:}
$$\ig_\O|\sum_{i,j}(\D^{-1}\dpr_i\dpr_j)(u_iu_j)|^qd\V\x\le
C^L_q\ig_\O|\V u|^{2q}d\V\x\ ,\quad 1<q<\io\Eq(2.3)$$
which we shall denote (CZ): here $\O$ is the torus of side $L$ and
$C^L_q$ is a suitable constant, [St93].
\*
 And finally
\*

\0(H) {\it H\"ol\-der inequality}:
$$\left|\ig f_1 f_2\ldots f_n \right|\le \prod_{i=1}^n
\left(\ig|f_i|^{p_i}\right)^\fra1{p_i},\qquad \sum_{i=1}^n \fra1{p_i}=1
 \Eq(2.4)$$
which we shall denote (H): the integrals are performed over an
arbitrary domain with respect to an arbitrary measure (of course the
same for all integrals).
\*

\0{\bf Remark:} The (H) are a (simple) extension of the Schwartz\--\-H\"ol\-der
inequalities; the (P) is a simple inequality for $\a=1$, \cfr problem [13];
while (P), (S) (mainly in the cases $\a=\fra32$ or $q=6$) and (CZ) are
less elementary and we refer to the literature, footnote at p.43
[So63], [St93], [LL01] for their proofs.
\*

An important consequence of the inequalities is
\*

{\bf Proposition 1:} {\it Let $\V u$ be a Leray solution verifying
(therefore) the {\it a priori} bounds in \equ(1.2): $\ig_\O |\V u(\xx,t)|^2
d\xx\le E_0$ and $\ig_0^Tdt\ig_\O|\W\dpr\V u(\xx,t)|^2\,d\V x\le
E_0\n^{-1}$ then
$$\ig_0^T dt\ig_\O d\xx\,|\V u|^{10/3}+\ig_0^T dt\ig_\O d\V x\, |p|^{5/3}
\le C \n^{-1}E_0^{5/3}\Eq(2.5)$$
where $C$ can be chosen $C^S_{\fra{10}3}\,(1+C^L_{\fra53})$.}
\*

\0{\it proof:} Apply (S) with  $q=\fra{10}3$ and $a=1$:
$$
\ig_\O |\V u|^{\fra{10}3}d\V x
\le C^S_{\fra{10}3}
\left(\ig_\O(\Dpr\W u)^2\,d\V x\right)^1
\cdot\left(\ig_\O \V u^2\,d\V x\right)^{\fra53-1}\le
C^S_{\fra{10}3} E_0^{\fra23}\ig_\O|\W\dpr\V u|^2\Eq(2.6)$$
hence integrating over $t$ between $0$ and $T$ using also the second
{\it a priori} estimate, we find
$$\ig_0^Tdt\ig_\O |\V u|^{\fra{10}3}d\V x\le C^S_{\fra{10}3}
E_0^\fra23\ig_0^Tdt\ig_\O d\V x\,(\W\dpr\V u)^2 \le
C^S_{\fra{10}3}\n^{-1} E_0^{1+\fra23}\Eq(2.7)$$
while the (CZ) yields: $\ig_\O d\V x\,|p|^\fra53\le C^L_\fra53
\ig_\O d\V x\,|\V u|^\fra{10}3$ which, integrated over $t$ and
combined with \equ(2.7), gives the announced result.
\*

\0{\bf \S3. Pseudo Navier Stokes velocity--pressure pairs.
Scaling operators.}
\numsec=3\numfor=1\*

As already mentioned the CKN theory will not fully use that $\V u$
verifies the Navier--Stokes equation: in order to better realize this
(unpleasant) property it is convenient to define separately the only
properties of the Leray solutions that are really needed to develop
the theory, \ie to obtain an estimate of the fractal dimension of the
space--time singularities set $S_0$.  This leads to the following
notion
\*

\0{\bf Definition a:} {\it (pseudo NS velocity field): Let $t\to(\V
u(\cdot,t),p(\cdot,t))$ be a function with values in the space of zero
average square integrable ``velocity'' and ``pressure'' fields on
$\O$. Suppose that for each $\f\in C^\io(\O\times(0,T])$ with
$\f(\xx,t)$ vanishing for $t$ near zero the following properties
hold. For each $T<\io$ and $s\le T$:
$$\eqalign{
(a)\quad&\ig_\O \V u\,d\xx =\V0, \qquad \V\dpr\cdot\V u=\V0,\qquad
p=-\sum_{i,j}\dpr_i\dpr_j
\D^{-1}(u_i u_j)\cr
(b)\quad&\ig_0^T dt\ig_\O d\V x\,
|\V u|^{10/3}+\ig_0^T dt\ig_\O d\V x\, |p|^{5/3}<\io\cr
(c)\quad&{1\over2}\ig_\O d\V x\,|\V u(\V x,s)|^2\f(\V x,s)+\n\ig_{t\le
s}\ig_\O\f(x,t)|\V\dpr\W u|^2d\V xd t\le\cr
&\le\ig_{t\le s}\ig_\O\Big[{1\over2}(\f_t+\n\,\D\f)|\V u|^2+
\fra12|\V u|^2\W u\cdot\W\dpr\f+p\,\W u\cdot\W\dpr\f
\Big]\,d\V x\, dt\cr}\Eq(3.1)$$
Then we shall say that the pair $(\V u,p)$ is a {\sl pseudo NS}
velocity and pressure pair. The {\sl singularity set} in the time
interval $[0,T]$ of $(\V u,p)$
will be defined as the set $S_0$ of the points
$(\xx,t)\in\O\times[0,T]$ that do not admit a vicinity $U$ where $|\V
u|$ is bounded.}\annota{4}{\nota Here we mean bounded outside a
set of zero measure in $U$ or, as one says, {\it essentially bounded}
because it is clear that, being $\V u,p$ in $L_2(\O)$, they are
defined up to a set of zero measure and it would not make sense to ask
that they are bounded everywhere without specifying which realization
of the functions we take.}
\*

The name given to the set $S_0$ is justified by a general result on
the theory of NS equations which shows that if a Leray's solution of
the NS equations is essentially bounded in a neighborhood of a space
time point then it is $C^\io$ near such point.
\*

\0{\bf Proposition 2:} {\it
(velocity is unbounded near singularities): Let $\V u(\xx,t)$ be a
Leray's solution of the NS equation in $L_2$.  Given $t_0>0$ suppose
that $|\V u(\xx,t)|\le M$, $(\xx,t)\in U_\r (\xx_0,t_0)\equiv$ sphere
of radius $\r$ ($\r<t_0$) around $(\xx_0,t_0)$, for some $M<\io$: then
$\V u\in C^\io(U_{\r/2}(\xx_0,t_0))$.}
\*

\0{\it Remarks:}
\\
\0(1) This means that the only way
a singularity can manifest itself, in a Leray solution of the NS
equations, is through a divergence of the velocity field itself. For
instance it is impossible to have a singular derivative having the
velocity itself unbounded. Hence, if $d\ge3$ velocity discontinuities
are impossible (and even less so shock waves), for instance. Naturally
if $\V u(\xx,t)$ is modified on a set of points $(\xx,t)$ with zero
measure it remains a weak solution (because the Fourier transform, in
terms of which the notion of weak solution is defined, does not
change), hence the condition $|\V u(\xx,t)|\le M$ {\it for each}
$(\xx,t)\in U_\l (\xx_0,t_0)$ can be replaced by the condition: {\it
for almost all} $(\xx,t)\in U_\l (\xx_0,t_0)$.

\0(2) The above result is not strong enough
to overcome the difficulties of a local theory of regularity of the
Leray weak solutions. Therefore one looks for other results of the
same type and it would be desirable to have results concerning
regularity implied by \ap informations on the vorticity. We have
already seen that bounded total vorticity implies regularity: however
it is very difficult to go really beyond; hence it is interesting to
note that also other properties of the vorticity may imply
regularity. A striking result in this direction, although insufficient
for concluding regularity (if true at all) of Leray weak solutions, is
in [CF93].

\0(3) For a proof of the above (Serrin's) theorem see [Ga02], proposition
IV in section 3.3.
\*

The remaining part of this section will concern the general properties
of the pseudo NS pairs and their regularity at a given point $(\V
x,t)$: it will not have more to do with the velocity and pressure
fields that solve the Navier--Stokes equations. It is indeed easy to
convince oneself that the \equ(3.1), in spite of the arbitrariness of
$\f$, {\it are not equivalent}, not even formally, to the
Navier--Stokes equations, and they pose far less severe on $\V u,p$
restrictions. We should not be surprised, therefore, if it turned out
possible to exhibit pseudo NS pairs that really have singularities on
``large sets'' of space--time. In a way it is already surprising that
the pseudo NS fields verify the regularity properties discussed below.

The analysis of the latter properties (of pseudo NS fields) is based
on the mutual relations between certain quantities that we shall
call ``{\it dimensionless operators}'' relative to the space--time
point $(\V x_0,t_0)$
\*

\0{\bf Definition b:} {\it (dimensionless ``operators'' for NS)
Let $(\V x_0,t_0)\in \O\times(0,\io)$ and consider the sets\annota{5}{If $r\ge
L/2$ this is interpreted as $B_r\=\O$. If $r^2\n^{-1}>t_0$ then
$\D_r(t_0)$ is interpreted as $0<t<t_0+r^2\n^{-1}$.}
$$\eqalign{
\D_r(t_0)=&\{ t|\,|t-t_0|<r^2\n^{-1}\}\cr
B_r(\V x_0)=&\{
\V\x\,|\, |\V\x-\V x_0|<r\}\=B_r\cr
Q_r(\V x_0,t_0)=&\{
(\V\x,\th)\,|\,|\V\x-\V x_0|<r,\,|\th-t_0|<r^2\n^{-1} \}\cr
Q_r(\V x_0,t_0)=&\D_r(t_0)\times B_r(\V x_0)\=Q_r\cr}\Eq(3.2)$$
define:
\*

\0{(i) } ``dimensionless kinetic energy operator'' on scale $r$:
$$A(r)=\fra1{\n^2 r}\sup_{|t-t_0|\le\n^{-1}r^2}\ig_{B_r}
|\V u(\V\x,t)|^2\,d\V \x\Eq(3.3)$$
and we say that the dimension of $A$ is $1$ : this refers to the factor
$r^{-1}$ that is used to make the integral dimensionless.

\0{(ii) }  ``local Reynolds number'' averaged on scale $r$:
$$\d(r)={1\over \n r}\ig_{Q_r}d\th d\V\x\,
|\W\dpr\,\V u|^2\Eq(3.4)$$
and we say that the dimension of $\d$ is $1$ : this refers in general
to the power $-\a$ to which $r$ has to be raised so that an
expression becomes dimensionless: in this case $\a=1$.

\0{(iii) }  ``dimensionless energy flux'' on scale $r$:
$$G(r)={1\over \n^2r^2}\ig_{Q_r}d\th d\V\x\, |\V
u|^3\Eq(3.5)$$
The dimension of $G$ is $2$.

\0{(iv) }  ``dimensionless pressure power'' forces on scale
$r$:
$$J(r)={1\over \n^2r^2}\ig_{Q_r} d\V\x d\th\,|\V u|\,|p|\Eq(3.6)$$
The dimension of $J$ is $2$.

\0{(v) } ``dimensionless non locality'' on scale $r$:
$$K(r)={r^{-13/4}\over\n^{3/2}}\ig_{\D_r}
d\th\Big(\ig_{B_r}|p|\,d\V\x\Big)^{5/4}\Eq(3.7)$$
The dimension of $K$ is $13/4$.

\0{(vi) } ``dimensionless intensity'' on  scale $r$:
$$S(r)=\n^{-7/3}r^{-5/3}\ig_{Q_r}(|\V u|^{10/3}+|p|^{5/3}) d\th d\V\x
\Eq(3.8)$$
where the pressure is always defined by the expression
$p=-\sum_{i,j=1}^3\dpr_i\dpr_j\D^{-1}(u_iu_j)$.  The dimension of $S$
is $5/3$.  }
\*

\0{\bf Remarks:} (1) The $A(r),\ldots$ {\it are not} operators in
the common sense of functional analysis. Their name is due to their
analogy with the quantities that appear in problems that are studied
with the methods of the ``renormalization group'' (which, also, are
not operators in the common sense of the words). Perhaps a more
appropriate name could be ``dimensionless observables'': but we shall
call them operators to stress the analogy of what follows with the
methods of the renormalization group.

(2) The $A(r),G(r),J(r),K(r),S(r)$ are in fact estimates of the
quantities that their name evokes. We omit the qualifier ``estimate''
when referring to them for brevity.

(3) The interest of (i)$\div$(iv) becomes manifest if we note that
the energy inequality \equ(3.1) can be expressed in terms of such
quantities if $\f$ is suitably chosen. Indeed let
$$\f=\chi(\V x,t)\,
\fra{\exp-\big(\fra{(\V x-\V x_0)^2}{4(\n(t_0-t)+2r^2)}\big)}{(4\p\n
(t-t_0)+8\p r^2)^{{3/2}} }\Eq(3.9)$$
where $\chi(\V x,t)$ is $C^\io$ and has value $1$ if $(\V x,t)\in
Q_{r/2}$ and $0$ if $(\V x,t)\not\in Q_r$. Then there exists a
constant $C>0$ such that
$$\eqalign{
&|\f|<\fra{C}{r^3},\quad|\Dpr\f|<\fra{C}{r^4},
\qquad|\dpr_t\f+\n\D\f|< \fra{C}{\n^{-1}r^5},\qquad {\rm everywhere}\cr
&|\f|> \fra1{C r^3},\kern5.cm {\rm if}\quad
(\V x,t)\in Q_{r/2}\cr}\Eq(3.10)$$
Hence \equ(3.1) implies
$$\fra{\n^2}{C r^2}\big( A({\txt\fra{r}2})+\d({\txt\fra{r}2})\big)\le
C\,\big(\fra1{\n^{-1} r^5}\ig_{Q_r}|\V u|^2+\fra1{r^4}\ig_{Q_r}|\V u|^3
+\fra1{r^4}\ig_{Q_r}|\V u||p|\big)\Eq(3.11)$$
and, since $\ig_{Q_r} |\V u|^2\le C \,(\ig_{Q_r}|\V
u|^3)^{2/3}(\n^{-1}r^5)^{1/3}$ with a suitable $C$, it follows that
for some $\tilde C$
$$A({\txt\fra{r}2})+\d({\txt\fra{r}2})\le \tilde C\ \big(G(r)^{2/3}+
G(r)+J(r)\big)\Eq(3.12)$$
(4) Note that the operator $\d(r)$ is an average of the 
``local Reynolds' number'' $r\ig_{\D_r} d\V\x |\W\dpr\V u|^2$.

(5) The operator (v) appears if one tries to bound $J(\fra{r}2)$ in
terms of $A(r)+\d(r)$: such an estimate is indeed possible and it will
lead to the {\it local Scheffer theorem} discussed in the next section.
\*

\0{\bf \S4. The theorems of Scheffer and of Caffarelli--Kohn--Nirenberg.}
\numsec=4\numfor=1\*

We can state the strongest results known (in general and to date)
about the regularity of the weak solutions of Navier Stokes equations
(which however hold also for the pseudo Navier Stokes
velocity--pressure pairs).
\*

\0{\bf Theorem I:} {\it (upper
bound on the dimension of the sporadic set of singular times for NS,
(``Scheffer's theorem'')): There are two constants $\e_s,C>0$ such
that if $G(r)+J(r)+K(r)<\e_s$ for a certain value of $r$, then $\V u$
is bounded in $Q_{\fra{r}2}(\V x_0,t_0)$:
$$|\V u(\V x,t)|\le C \fra{\e_s^{1/3}}r,\quad (\V x,t)\in
Q_{\fra{r}2}(\V x_0,t_0),\quad {\rm almost\ everywhere}\Eq(4.1)$$
having set $\n=1$.}
\*

\0{\it Remarks:} (1) \cfr problems
[5]$\div$[11] for a guide to the proof.

\0(2) This theorem can be conveniently combined, for the purpose of
checking its hypotheses, with the inequality: $J(r)+G(r)+K(r)\le
C\,\big( S(r)^{9/10}+S(r)^{3/4}\big)$, which follows immediately from
inequality (H) and from the definitions of the operators, with a
suitable $C$.

\0(3) In other words {\it if the operator $S(r)$ is small enough then
$(\xx_0,t_0)$ is a regular point.}

\0(4) This will imply that the fractal dimension of the space--time
singularities set is $\le5/3$. In fact, see section 5 below, an \ap
estimate on the global value of an operator with dimension $\a$
implies that the Hausdorff' measure of the set of points around which
the operator is large does not exceed $\a$; here the operator $S(r)$
has dimension $5/3$ and therefore together with the \ap bound \equ(2.5)
it yields and estimate $\le 5/3$ for the Hausdorff dimension of the
singularity set.  This also justifies the introduction of the operator
$S(r)$.
\*

It is easy, in terms of the just defined operators, to illustrate the
strategy of the proof of the following theorem which will immediately
imply, via a classical argument reproduced in section 5 below, that
the fractal dimension of the space time singularities set $S_0$ for a
pseudo NS field is $\le1$ and that its $1$--measure of Hausdorff
$\m_1(S_0)$ vanishes.
\*

\0{\bf Theorem II:} {\it (sufficient condition for local regularity
space-time (``CKN theorem'')) There is a
constant $\e_{ckn}$ such that if $(\V u,p)$ is a pseudo NS pair of
velocity and pressure fields and
$$\limsup_{r\to0}\fra1{\n r}\ig_{Q_r(\V x_0,t_0)}|\W\dpr\V u(\V
x',t')|^2\,d\V x' dt'\=\limsup_{r\to0}\,\d(r)\,<\,\e_{ckn}\Eq(4.2)$$
then $\V u(\V x',t'), \,p(\xx',t')$ are $C^\io$ in the vicinity of $(\V
x_0,t_0)$.}\annota{6}{\nota This means that near $(\V x,t)$ the
functions $\V u(\V x',t'), p(\V x',t')$ coincide with $C^\io$
functions apart form a set of zero measure (recall that the pseudo NS
fields are defined as fields in $L_2(\O)$).}
\*

For fixed $(\xx_0,t_0)$, consider the ``sequence of length scales'':
$r_n\=L 2^n$, with $n=0,-1,-2,\ldots$. We shall set $\a_n\=A(r_n)$,
$\k_n=K_n^{8/5}$, $j_n=J_n$, $g_n=G_n^{2/3}$, $\d_n=\d(r_n)$ which is
a natural definition as it will shortly appear.  And define $\V
X_n\=(\a_n,\k_n,j_n,g_n)\in R_+^4$.
Then the proof of this theorem is based on a bound that allows us to
estimate the size of $\V X_n$, defined as the sum of its components, in
terms of the size of $\V X_{n+p}$  provided the Reynolds
number $\d_{n+p}$ on scale $n+p$ is $\le \d$.

The inequality will have the form (if $p>0$ and $0<\d<1$)
$$\V X_n\le \BB_p(\V X_{n+p};\d)\Eq(4.3)$$
where $\BB_p(\cdot;\d)$ is a map of the whole $R_+^4$ into itself and
the inequality has to be understood ``component wise'', \ie in the sense
that each component of the l.h.s. is bounded by the corresponding
component of the r.h.s. We call $|\V X|$ the sum of the components of
$\V X\in R^4_+$.

The map $\BB_p(\cdot;\d)$, which to some readers will appear as strongly
related to the ``{\it beta function}'' for the ``running couplings'' of
the ``renormalization group approaches'',\annota{7}{\nota Indeed it
relates properties of operators on a scale to those on a different
scale. Note, however, that the couplings on scale $n$, \ie the
components of $\V X_n$, provide information on those of $X_{n+p}$ rather
than on those of $\V X_{n-p}$ as usual in the renormalization group
methods, see [BG95].} will enjoy the following property
\*

\0{\bf Proposition 3:} {\it Suppose that $p$ is large enough; given $\r>0$
there exists $\d_p(\r)>0$ such that if $\d<\d_p(\r)$ then the iterates
of the map $\BB_p(\cdot;\d)$ contract any given ball in $R^4_+$,
within a finite number of iterations, into the ball of radius $\r$:
\ie $|\BB_p^k(\V X;\d)|<\r$ for all large $k$'s.}
\*

Assuming the above proposition the main theorem II follows:
\*

\0{\it proof}: Let $\r=\e_s$, \cfr theorem I, and let
$p$ be so large that the above proposition holds. We set
$\e_{ckn}=\d_p(\e_s)$ and it will be, by the assumption \equ(4.2), that
$\d_n<\e_{ckn}$ for all $n\le n_0$ for a suitable $n_0$ (recall that
the scale labels $n$ are negative).

Therefore it follows that $|\BB^k_p(\V X_{n_0};\e_{ckn})|<\e_s$ for some
$k$. Therefore by the theorem I we conclude that $(\xx_0,t_0)$ is a
regularity point.
\*

\0{\it Proof that the renormalization map contracts.}
\*
Proposition 3 follows immediately from the following general ``Sobolev
inequalities''
\*

(1) {\it ``Kinematic inequalities'':} \ie
inequalities depending only on the fact that $\V u$ is a divergence
zero, average zero and is in $L_2(\O)$ and $p=-\D^{-1}(\W\dpr\,\V
u)^2$
$$\eqalign{ J_n&\le
C\,(2^{-p/5}A^{1/5}_{n+p}G^{1/5}_nK^{4/5}_{n+p}+2^{2p}A^{1/2}_{n+p}\d_{n+p})
\cr K_n&\le C\,(2^{-p/2}K_{n+p}+2^{5p/4}A^{5/8}_{n+p}\d^{5/8}_{n+p})\cr
G_n^{2/3}&\le C\,(2^{-2p}A_{n+p}+2^{2p}A^{1/2}_{n+p}\d^{1/2}_{n+p})\cr}
\Eq(4.4)$$
where $C$ denotes a suitable constant ({\it independent on the
particular pseudo NS field}).  The proof of the inequalities
\equ(4.4) is not difficult, assuming the (S,H,CZ,P) inequalities
above, and it is illustrated in the problems [1], [2], [3].

(2) {\it ``Dynamical inequality'':} \ie an inequality based on the
energy inequality (c) in \equ(3.1) which implies, quite easily, the
following ``{\it dynamic inequality}''\annota{8}{\nota We call
it ``dynamic'' because it follows from the energy inequality, \ie from
the equations of motion.}
$$A_n\le
C\,(2^pG^{2/3}_{n+p}+2^pA_{n+p}\d_{n+p}+2^pJ_{n+p})\Eq(4.5)$$
whose proof is illustrated in problem [4].
\*

\0{\it proof of proposition:} Assume the above inequalities
\equ(4.4), \equ(4.5) and setting $\a_n=A_n, \k_n=K_n^{8/5},j_n=J_n,
g_n=G_n^{2/3}$, $\d_{n+p}=\d$ and, as above, $\V
X_n=(\a_n,\k_n,j_n,g_n)$. The r.h.s. of the inequalities defines the map
$\BB_p(\V X;\d)$.

If one stares long enough at them one realizes that the contraction
property of the proposition is an immediate consequence of
\\
(1) The exponents to which $\e=2^{-p}$ is raised in the various terms
are either positive or not; in the latter cases the inverse power of
$\e$ is always appearing multiplied by a power of $\d_{n+p}$ which we
can take so small to compensate for the size of $\e$ to any negative
power, {\it except in the one case corresponding to the last term in}
\equ(4.5) where we see $\e^{-1}$ without any compensating $\d_{n+p}$.
\\
(2) Furthermore the sum of the powers of the components of $\V X_n$ in
each term of the inequalities is {\it always $\le1$}: this means that
the inequalities are ``almost linear'' and a linear map that ``bounds''
$\BB_p$ exists and it is described by a matrix with small entries {\it
except one off--diagonal element}. The iterates of the matrix therefore
contract unless the large matrix element ``ill placed'' in the matrix:
and one easily sees that it is not.

A formal argument can be devised in many ways: we present one in
which several choices appear that are quite arbitrary and that the
reader can replace with alternatives. In a way one should really try to
see why a formal argument is not necessary.
\*
The relation \equ(4.5) can be ``iterated'' by using the
expressions \equ(4.4) for $G_{n+p},J_{n+p}$ and then the first of
\equ(4.4) to express $G_{n+p}^{1/5}$ in terms of $A_{n+2p}$
with $n$ replaced by $n+p$:

$$\eqalign{
\a_n\,\le\, C\,& (2^{-p} \a_{n+2p} +2^{3p}
\d_{n+2p}^{1/2}\a_{n+2p}^{1/2}+\cr
&+2^{p/5}(\a_{n+2p}\k_{n+2p})^{1/2}
+2^{7p/5}\d_{n+2p}\a^{7/{20}}_{n+2p}\k_{n+2p}^{1/2}+\cr
&+2^{3p}\d_{n+2p}\a_{n+2p})\cr}\Eq(4.6)$$

It is convenient to take advantage of the simple inequalities
$(ab)^{\fra12}\le z a + z^{-1} b$ and $a^x\le 1+a$ for
$a,b,z,x>0, \, x\le1$. 

The \equ(4.6) can be turned into a relation between $\a_n$ and
$\a_{n+p},\k_{n+p}$ by replacing $p$ by $\fra12p$. Furthermore, in the
relation between $\a_n$ and $\a_{n+p},\k_{n+p}$ obtained after the
latter replacement, we choose $z=2^{-p/5}$ to disentangle
$2^{p/10}(\a_{n+p}\k_{n+p})^{1/2}$ we obtain recurrent (generous)
estimates for $\a_n,\k_2$

$$\eqalign{
\a_n\,\le\,& C\,( 2^{-p/10} \a_{n+p} +2^{3p/10}\k_{n+p} + \x^\a_{n+p})
\cr
\k_n\,\le\,& C\,(2^{-4p/5}\k_{n+p}+\x^\k_{n+p})\cr
\x^\a_{n+p}\,\defi\,& 2^{3p}\d_{n+p}(\a_{n+p}+\k_{n+p}+1)\cr
\x^\k_{n+p}\,\defi\,& 2^{3p}\d_{n+p}\a_{n+p}\cr}\Eq(4.7)$$
We fix $p$ once and for all such that $2^{-p/10} C<\fra13$.

Then if $C 2^{3p}\d_n$ is small enough, \ie if $\d_n$ is small enough,
say for $\d_n<\lis\d$ {\it for all $|n|\ge \lis n$}, the matrix
$M=C\,\pmatrix{2^{-p/10} +2^{3p}\d_{n+p}& 2^{3p/10}+2^{3p}\d_{n+p}\cr
0& 2^{-4p/5}+2^{3p}\d_{n+p}\cr}$ will have the two eigenvalues
$<\fra12$ and iteration of \equ(4.6) will contract any ball in the
plane $\a,\k$ to the ball of radius $2\lis\d$.

If $\a_n,\k_n$ are bounded by a constant $\lis\d$ for all $|n|$ large
enough the \equ(4.4) show that also $g_n,j_n$ are going to be
eventually bounded proportionally to $\lis\d$.

Hence by imposing that $\d$ is so small that
$|X_n|=\a_n+\k_n+j_n+g_n<r$ $\r$ we see that proposition 3 holds
(hence theorem 2 as a consequence of theorem 1).
\*

\0{\bf \S5. Fractal
dimension of singularities of the Navier--Stokes equation, $d=3$.}
\numsec=5\numfor=1\*

Here we ask which could be the structure of the possible set of the
singularity points of the solutions of the Navier--Stokes equation in
$d=3$. The answer is an immediate consequence of theorem II and we
describe it here for completeness: the technique is a classic method
(Almgren) to link \ap estimates to fractal dimension estimates.

It has been shown already by Leray that the set of times at which a
singularity is possible has zero measure (on the time axis), see \S3.4
in [Ga02].

Obviously sets of zero measure can be quite structured and even large
in other senses.  One can think to the example of the Cantor
set which is non denumerable and obtained from an
interval $I$ by deleting an open concentric subinterval of length
$1/3$ that of $I$ and then repeating recursively this operation on
each of the remaining intervals (called $n$--th generation intervals
after $n$ steps); or one can think to the set of rational points which
is dense.
\*

\0{\it (A) Dimension and measure of Hausdorff.}
\*

An interesting geometric characteristic of the size of a set is given
by the Hausdorff dimension and by the Hausdorff measure, \cfr
[DS60], p.174.
\*

\0{\bf Definition c:} {\it (Hausdorff $\a$--measure):
The Hausdorff $\a$-measure of a set $A$ contained in a metric space
$M$ is defined by considering for each $\d>0$ all coverings $\CC_\d$
of $A$ by closed sets $F$ with diameter $0<d(F)\le\d$ and setting
$$\m_\a(A)=\lim_{\d\to0}\,\inf_{\CC_\d}\,\sum_{F\in
\CC_\d}d(F)^\a\Eq(5.1)$$
}

\0{\bf Remarks:} (1) The limit over $\d$ exists because the quantity
$\inf_{\CC_\d}\ldots$ is monotonic nondecreasing.

\0(2) It is possible to show that the function defined
on the sets $A$ of $M$ by $A\to\m_\a(A)$ is completely additive
on the smallest family of sets containing all closed sets and
invariant with respect to the operations of complementation and
countable union (which is called the {\it $\s$-algebra $\Zi$ of the
{\it Borel sets} of $M$}), \cfr [DS60].
\*

One checks immediately that given $A\in \Zi$ there is $\a_c$ such that
$$
\m_\a(A)=\io\ {\rm if}\quad\a<\a_c,\qquad
\m_\a(A)=0 \ {\rm if}\ \a>\a_c\Eq(5.2)$$
and it is therefore natural to set up the following definition
\*
\0{\bf Definition d:} {\it (Hausdorff measure and Hausdorff dimension):
Given a set $A\subset R^d$ the quantity $\a_c$, \equ(5.2), is called
{\sl Hausdorff dimension} of $A$, while $\m_{\a_c}(A)$ is called {\sl
Hausdorff measure} of $A$.}
\*

It is not difficult to check that
\*
\0(1) Denumerable sets in $[0,1]$ have zero Hausdorff dimension and measure.

\0(2) Hausdorff dimension of
$n$-dimensional regular surfaces in $R^d$ is $n$ and, furthermore,
the Hausdorff measure of their Borel
subsets defines on the surface a measure $\m_{\a_c}$ that is equivalent
to the area measure $\m$: namely there is a $\r(x)$ such that
$\m_{\a_c}(dx)=\r(x)\,\m(dx)$.

\0(3) The Cantor set, defined also as the set of all numbers in $[0,1]$
which in the representation in base $3$ do not contain the digit $1$,
has
$$\a_c=\log_3 2\Eq(5.3)$$
as Hausdorff dimension.\annota{9}{
Indeed with $2^n$ disjoint segments with size $3^{-n}$, uniquely
determined (the $n$--th generation segments), one covers the whole set
$C$; hence
$$\m_{\a,\d}\defi\inf_{\CC_\d}\sum_{F\in
\CC_\d}d(F)^\a\le1\qquad{\rm if}
\quad\a=\a_0=\log_32$$
and $\m_{\a_0}(C)\le1$: \ie $\m_\a(C)=0$ if $\a>\a_0$. Furthermore,
\cfr problem [16] below, if $\a<\a_0$ one checks that the covering
$\CC^0$ realizing the smallest value of $\sum_{F\in \CC_\d}d(F)^\a$
with $\d=3^{-n}$ is precisely the just considered one consisting in
the $2^n$ intervals of length $3^{-n}$ of the $n$--th generation and
the value of the sum on such covering diverges for $n\to\io$.  Hence
$\m_\a(C)=\io$ if $\a<\a_0$ so that $\a_0\=\a_c$ and $\m_{\a_c}(C)=1$.}
\*

\0{\it(B) Hausdorff dimension of singular times in the Navier--Stokes
solutions ($d=3$).}
\*

We now attempt to estimate the Hausdorff dimension of the sets of
times $t\le T<\io$ at which appear singularities of a given weak
solution of Leray, \ie a solution of the type discussed in
\equ(1.1).  Here $T$ is an \ap arbitrarily prefixed time.

We need a key property of Leray's solutions, namely that if at time
$t_0$ it is $J_1(t_0)=L^{-1}\ig (\V\dpr \T u)^2 d\V x<\io$,
\ie if the Reynolds number $R(t_0)=J_1(t_0)^{1/2}/V_c\=V_1/V_c$ with
$V_c\defi \n L^{-1}$
is $<+\io$, then the solution stays regular in a time
interval $(t_0,t_0+\t]$ with (see proposition II in \S3.3 of [Ga02],
eq. (3.3.34)):
$$\t=F{T_c\over R(t_0)^4},\qquad T_c=\fra{L^2}\n\Eq(5.4)$$
{}From this it will follow, see below, that there are $A>0,\g>0$ such that if
$$\liminf_{\s\to0} \Big({\s\over T_c}\Big)^{\g}\ig_{t-\s}^t {d\th\over\s}
R^2(\th)< A\Eq(5.5)$$
then $\t>\s$ and the solution is regular in an interval that contains
$t$ so that the instant $t$ is an instant at which the solution is
regular. Here, as in the following, we could fix $\g=1/2$: but $\g$ is
left arbitrary in order to make clearer why the choice $\g=1/2$ is the
``best''.

We first show that, indeed, from \equ(5.5) we deduce the existence of a
sequence $\s_i\to0$ such that
$$\ig_{t-\s_i}^t{d\th\over\s_i}\,R^2(\th)<A\Big({\s_i\over T_c}
\Big)^{-\g}\Eq(5.6)$$
therefore, the l.h.s. being a time average, there must exist
$\th_{0i}\in(t-\s_i,t)$ such that
$$R^2(\th_{0i})< A \Big({\s_i\over T_c}\Big)^{-\g}\Eq(5.7)$$
and then the solution is regular in the interval
$(\th_{0i},\th_{0i}+\t_i)$ with length $\t_i$ at least
$$\t_i=FT_c {(\s_i/T_c)^{2\g}\over A^2}>\s_i\Eq(5.8)$$
{\it provided} $\g\le1/2$, and $\s_i$ is small enough and if $A$ is
small enough (if $\g=\fra12$ then this means $2A^2< F$).  Under these
conditions the size of the regularity interval is longer than $\s_i$
and {\it therefore it contains $t$ itself}.

It follows that, if  $t$ is in the set $S$ of the times
at which a singularity is present, it must be
$$\liminf_{\s\to0}\Big(
{\s\over T_c}\Big)^\g\ig_{t-\s}^t{d\th\over\s}\,R^2(\th)\ge A
\qquad{\rm if}\ t\in S\Eq(5.9)$$
\ie every singularity point is covered by a family of infinitely many
intervals $F$ with diameters $\s$ {\it arbitrarily small} and
satisfying
$$\ig_{t-\s}^t d\th\,R^2(\th)\ge {A\over 2}\s\Big({\s\over
T_c}\Big)^{-\g}\Eq(5.10)$$
{}From Vitali's covering theorem (\cfr problem [19]) it follows that, given
$\d>0$, one can find a denumerable family of intervals
$F_1,F_2,\ldots$, with $F_i=(t_i-\s_i,t_i)$, pairwise disjoint and
verifying the \equ(5.10) and $\s_i<\d/4$, such that the intervals
$5F_i\defi(t_i-7\s_i/2,t_i+5\s_i/2)$ (obtained by dilating the
intervals $F_i$ by a factor $5$ about their center) cover $S$
$$S\subset\cup_i\,5F_i\Eq(5.11)$$
Consider therefore the covering $\CC$ generated by the sets
$5F_i$ and compute the sum in \equ(5.1) with $\a=1-\g$:
$$\eqalign{
&\sum_i(5\s_i)\Big({5\s_i\over T_c}\Big)^{-\g}
=5^{1-\g}\sum_i\s_i\Big({\s_i\over T_c}\Big)^{-\g}<\cr
&<{2\,5^{1-\g}\over A\sqrt{T_c}}
\sum_i\ig_{F_i}d\th\,R^2(\th)\le {2 \,5^{1-\g}\over A\sqrt{T_c}}
\ig_0^T d\th\,R^2(\th) <\io\cr}\Eq(5.12)$$
where we have made use of the \ap estimates on vorticity \equ(1.2) and
we must recall that $\g\le1/2$ is a necessary condition in order that
what has been derived be valid (\cfr comment to \equ(5.8)).

Hence it is clear that for each $\a\ge1/2$ it is $\m_\a(S)<\io$ (pick,
in fact, $\a=1-\g$, with $\g\le1/2$) hence the Hausdorff dimension of
$S$ is $\a_c\le1/2$. Obviously the choice that gives the best
regularity result (with the informations that we gathered) is
precisely $\g=1/2$.

Moreover one can check that $\m_{1/2}(S)=0$: indeed we know that $S$
has zero measure, hence there is an open set $G\supset S$ with measure
smaller than a prefixed $\e$. And we can choose the intervals $F_i$
considered above so that they also verify $F_i\subset G$: hence we can
replace the integral in the right hand side of \equ(5.12) with the
integral over $G$ hence, since the integrand is summable, we shall find
that the value of the integral can be supposed as small as wished, so
that $\m_{1/2}(S)=0$.

\*
\0{\it(C) Hausdorff dimension in space--time of the solutions of NS,
($d=3$).}
\*

The problem of which is the Hausdorff dimension of the points $(\V
x,t)\in\O\times[0,T]$ which are singularity points for the Leray's
solutions is quite different.

Indeed, \ap, it could even happen that, at one of the times $t\in S$
where the solution has a singularity as a function of time, {\it all}
points $(\V x,t)$, with $\V x\in \O$, are singularity points and
therefore the set $S_0$ of the singularity points thought of as a set
in space--time could have dimension $3$ (and perhaps even $3.5$ if we
take into account the dimension of the singular times discussed in (B)
above).
\*

From theorem II we know that if

$$\limsup_{r\to0}r^{-1}\ig_{t-r^2/2\n}^{t+r^2/2\n}\ig_{S(\xx,r)}
{d\th\over\n}\,d\V\x\,(\V\dpr\T u)^2<\e\Eq(5.13)$$
then regularity at the point $(\V x,t)$ follows.

It follows that the set $S_0$ of the singularity points in space--time
can be covered by sets $C_r=S(\V
x,r)\times(t-r^2\n^{-1},t+\fra12r^2\n^{-1}]$ with $r$ arbitrarily
small and such that
$$\fra1{r\n}\ig_{t-\fra{r^2}{2\n}}^{t+\fra{r^2}{2\n}} d\th \ig_{S(\V
x,r)} d\V x\,(\V\dpr\T u)^2>\e\Eq(5.14)$$
which is the negation of the property in  \equ(5.13).

Again by a covering theorem of Vitali (\cfr problems [16],[17]), we
can find a family $F_i$ of sets of the form $F_i=S(\V
x_i,r_i)\times(t_i-\fra{r_i^2}{2\n},t_i +\fra{r_i^2}{2\n}]$ pairwise
disjoint and such that the sets $6F_i$= set of points $(\V x',t')$ at
distance $\le 6r_i$ from the points of $F_i$ covers the singularity
set $S_0$.\annota{{10}}{\nota Here the constant $5$, as well as the
other numerical constants that we meet below like $5,6,18$ have no
importance for our purposes and are just simple constants for which
the estimates work.} One can then estimate the sum in \equ(5.1) for
such a covering, by using that the sets $F_i$ are pairwise disjoint
and that $5F_i$ has diameter, if $\max r_i$ is small enough, not
larger than $18 r_i$:
$$\sum_i (18r_i)\le{36\over \n\e}\sum_i\ig_{F_i}(\V\dpr\T u)^2d\V\x
dt\le {36\over\n \e}\ig_0^T\ig_\O(\V\dpr\T u)^2d\V\x dt<\io
\Eq(5.15)$$
\ie the $1$-measure of  Hausdorff $\m_1(S_0)$ would be $<\io$ hence
the Hausdorff dimension of  $S_0$ would be $\le 1$.

Since $S_0$ has zero measure, being contained in $\O\times S$ where
$S$ is the set of times at which a singularity occurs somewhere, see
\equ(5.9), it follows (still from the covering theorems) that in
fact it is possible to choose the sets $F_i$ so that their union $U$
is contained into an open set $G$ which differs from $S_0$ by a set of
measure that exceeds by as little as desired that of $S_0$ (which is
zero); one follows the same method used above in the analysis of the
time--singularity.  Hence we can replace the last integral in
\equ(5.15) with an integral extended to the union $U$ of the $F_i's$:
the latter integral can be made as small as wished by letting the
measure of $G$ to $0$.  It follows that not only the Hausdorff
dimension of $S_0$ is $\le1$, but also the $\m_1(S_0)=0$.
\*

\0{\bf Remarks:} (1) In this way we exclude that the set $S_0$
of the space--time singularities contains a regular curve:
singularities, {\it if existent}, cannot move along trajectories (like
flow lines) otherwise the dimension of $S$ would be $1>1/2$) nor they
can distribute, at fixed time, along lines and, hence, in a sense they
must appear isolates and immediately disappear (always assuming their
real existence).

\0(2) A conjecture (much debated and that I favor) that is
behind all our discussions is that {\it if the initial datum $\V u^0$
is in $C^\io(\O)$ then there exists a solution to the
Navier Stokes equation that is of class $C^\io$ in $(\xx,t)$}'', \ie
$S_0=\emptyset$!
The problem is, still, open: counterexamples to the conjecture are not
known (\ie singular Leray's solutions with initial data and external
force of class $C^\io$) but the matter is much debated and different
alternative conjectures are possible (\cfr [PS87]).

\0(3) In this respect one should keep in mind that if $d\ge4$ it is possible
to show that {\it not all} smooth initial data evolve into regular
solutions: counterexamples to smoothness can indeed be constructed,
\cfr [Sc77]).
\*

\def\tit{\nota Dimensional bounds of the theory CKN}
\0{\bf Problems. The dimensional bounds of the  CKN theory.}
\*

{\nota In the following problems we shall set $\n=1$, with no loss of
generality, thus fixing the units so that time is a square length. The
symbols $(\V u,p)$ will denote a pseudo NS field, according to
definition 1. Moreover, for notational simplicity, we shall set
$A_\r\=A(\r)$, $G_\r\=G(\r),\ldots$, and sometimes we shall write
$A_{r_n}, G_{r_n}\ldots$ as $A_n,\ldots$ with an abuse that should not
generate ambiguities. The validity of the \equ(3.1) for Leray's
solution is checked in problem [15], at the end of the problems
section, to stress that the theorems of Scheffer and CKN concern
pseudo NS velocity--pressure fields: however it is independent of the
first $14$ problems. There will many constants that we generically
denote $C$: they are not the same but one should think that they
increase monotnically by a finite amount at each inequality. The
integration elements like $d\xx$ and $dt$ are often omitted to
simplify the notations and they should be easily understood from the
integration domains.
\*

\0{\bf[1]:} Let $\r=r_{n+p}$ and $r=r_n$, with $r_n=L 2^n$, \cfr
lines following \equ(4.2), and apply (S),\equ(2.2), with $q=3$ and
$a=\fra34$, to the field $\V u$, at $t$ fixed in $\D_r$ and using
definition 2 deduce
$$\eqalign{
\ig_{B_r} |\V u|^3d\V x&\le C^S_3 [
\Big(\ig_{B_r}|\Dpr\W u|^2\,d\V x\Big)^{\fra34}
\Big(\ig_{B_r}|\V u|^2\,d\V
x\Big)^{\fra34}+{r^{-3/2}}\Big(\ig_{B_r}|\V
u|^2\Big)^{3/2}]\le\cr
&\le C^S_3[ \r^{3/4} A_\r^{3/4}
\Big(\ig_{B_r}|\W\dpr\V u|^2d\V
x\Big)^{3/4}+
r^{-3/2}\Big(\ig_{B_r}|\V u|^2\Big)^{3/2}]\cr}$$
\0Infer from the above the third of  \equ(4.4). (\Idea Let
$\lis{|\V u|^2_\r}$ be the average of $\V u^2$ on the ball $B_\r$; apply
the inequality (P), with $\a=1$, to show that there is $C>0$ such that
$$\eqalign{
&\ig_{B_r}d\xx\,|\V u|^2\le\Big(\ig_{B_\r}d\V x\,
\Big||\V u|^2-\lis{|\V u|^2_\r}
\Big|\Big)+\lis{|\V u|^2_\r}\ig_{B_r}d\V x\,\le\cr
&\le C \r\ig_{B_\r}d\V x\,|\V u||\W\dpr\V u|+C
\Big(\fra{r}\r\Big)^3\ig_{B_\r}d\V x\,|\V u|^2\le
C \r^{3/2} A_\r^{1/2}\Big(\ig_{B_\r}d\V x\,|\W\dpr\V u|^2\Big)^{1/2}
+\cr
&+C\Big(\fra{r}\r\Big)^3\r A_\r\cr}$$
where the dependence from $t\in \D_r$ is not explicitly indicated;
hence
$$\ig_{B_r}d\V x\,|\V u|^3
\le C\,(r\r^{-1})^3 A_\r^{3/2}+C\,(\r^{3/4}+\r^{9/4}r^{-3/2})
A_\r^{3/4}\Big(\ig_{B_\r}d\xx\,|\W\dpr\V u|^2\Big)^{3/4}$$
then integrate both sides with respect to $t\in \D_r$ and apply (H) and
definition 2.)
\*

\0{\bf[2]:} Let $\f\le1$ be a non negative $C^\io$ function with
value $1$ if $|\V x|\le 3\r/4$ and $0$ if $|\V x|>4\r/5$; we suppose
that it has the ``scaling'' form $\f=\f_1(\xx/\r$ with $\f_1\ge0$
a $C^\io$ function fixed once and for all. Let $B_\r$ be the ball
centered at $\xx$ with radius $\r$; and note that, if $\r=r_{n+p}$ and
$r=r_n$, pressure can be written, at each time (without explicitly
exhibiting the time dependence), as $p(\xx)=p'(\xx)+p''(\xx)$ with
$$\eqalign{
p'(\xx)=&\fra1{4\p}\ig_{B_\r}
\fra1{|\V x-\V y|} p(y) \D\f(\V y)\,d\V y+\fra1{2\p}
\ig_{B_\r}\fra{\V x-\V y}{|\V x-\V y|^3}\cdot\V\dpr\f(\V y)\,p(\V y)\,d\yy\cr
p''(\xx)=&\fra1{4\p}\ig_{B_\r}\fra1{|\V x-\V y|}\f(\V y)\,(\W\dpr\V
u(\V y))\cdot(\V\dpr\W u(\V y))\,d\V y\cr}$$
if $|\V x|<3\r/4$; and also $|p'(\xx)|\le C\r^{-3}\ig_{B_\r}d\yy\,
|p(\yy)|$ and all functions are evaluated at a fixed $t\in \D_r$.
Deduce from this remark the first of the
\equ(4.4). (\Idea First note the identity $p=-(4\p)^{-1}
\ig_{B_\r}|\xx-\yy|^{-1}\,\D\,(\f\,p)$ for $\xx\in B_r$ because
if $\xx\in B_{3\r/4}$ it is $\f\,p\=\,p$. Then note the identity
$\D\,(\f\,p)= p\,\D\,\f+2\V\dpr\,p\cdot\V\dpr \f+\f\,\D\,p$ and since
$\D\,p=-\V\dpr\cdot(\W u\cdot\W\dpr \,\V u)=-(\V\dpr\, \W
u)\cdot(\W\dpr\,\V u)$: the second of the latter relations generates
$p''$ while $p\D\,\f$ combines with the contribution from
$2\V\dpr\,p\cdot\V\dpr \f$, after integrating the latter by parts, and
generates the two contributions to $p'$.
\\
{}From the expression for $p''$ we see that
$$\eqalign{
\ig_{B_r}d\xx\,|p''(\xx)|^2&\le\ig_{B_\r\times B_\r}d\yy\,d\yy'\,
|\W\dpr\V u(\V y)|^2|\W\dpr\V u(\V y')|^2
\ig_{B_r}d\V x\fra1{|\V x-\V y||\V x-\V y'|}\le \cr
&\le C\r(\ig_{B_\r}d\yy\,|\W\dpr\V u(\yy)|^2)^2\cr}\eqno(!)$$
The part with $p'$ is more interesting: since its expression above
contains inside the integral kernels apparently singular at $\xx=\yy$
like $|\xx-\yy|^{-1}\,\D\,\f$ and $|\xx-\yy|^{-1}\,\Dpr\,\f$ one
remarks that, in fact, there is no singularity because the derivatives
of $\f$ vanish if $\yy\in B_{3\r/4}$ (where $\f\=1$). Hence 
$|\xx-\yy|^{-k}$ can be bounded ``dimensionally'' by $\r^{-k}$ in the
whole region $B_\r/B_{3\r/4}$ for all $k\ge0$ (this remark also motivates
why one should think $p$ as sum of $p'$ and $p''$).

Thus replacing the (apparently) singular kernels with their
dimensional bounds  we get
$$\ig_{B_r}d\xx\,|\V u||p'|\le
\fra{C}{\r^3}\Big(\ig_{B_r}d\xx\,|\V u|\Big)\cdot\Big(\ig_{B_\r}
d\xx\,|p|\Big)$$
which can be bounded by using inequality (H) as
$$\eqalign{
&\le \fra{C}{\r^3}\Big(\ig_{B_r}d\xx\,|\V u|^{2/5}\cdot|\V
u|^{3/5}\cdot1\Big)\cdot\Big(\ig_{B_\r}d\xx\,|p|\Big)\le\cr
&\le \fra{C}{\r^3}\Big(\ig_{B_r}d\xx\,|\V u|^{2}\Big)^{1/5}\cdot\Big(
\ig_{B_r}d\xx\,|\V
u|^{3}\Big)^{1/5}\,(r^3)^{3/5}\cdot\,\ig_{B_\r}d\xx\,|p|\le\cr
&\le \fra{C}{\r^3}\,(\r A_\r)^{1/5}\Big(\ig_{B_r}d\xx\,|\V
u|^{3}\Big)^{1/5}
\cdot\Big(\ig_{B_\r}d\xx\,|p|\Big)\cr}$$
where all functions depend on $\xx$ (and of course on $t$) and then,
integrating over $t\in\D_r$ and dividing by $r^2$ one finds, for a
suitable $C>0$:
$$\fra1{r^2}\ig_{Q_r}dtd\xx\,|\V u||p'|\le
C\,(\fra{r}\r)^{1/5}G_r^{1/5}K_\r^{4/5}A_\r^{1/5}$$
that is combined with $\ig_{B_r}d\xx|\V u||p''|\le (\ig_{B_r}d\xx\,|\V
u|^2)^{1/2}(\ig_{B_r}d\xx\,|p''|^2)^{1/2}$ which, integrating over
time, dividing by $\r^2$ and using inequality (!) for $\ig_{B_r}d\xx\,
|p''|^2$ yields: $r^{-2}\ig_{Q_r}dtd\xx\,|\V u||p''|\le C(\r r^{-1})^2
A_\r^{1/2}\d_\r$).
\*

\0{\bf[3]} In the context of the hint and notations for $p$ of the
preceding problem check that $\ig_{B_r}d\xx\,|p'|\le C
(r\r^{-1})^3\ig_{B_\r}d\xx|p|$. Integrate over $t$ the power $5/4$ of
this inequality, rendered adimensional by dividing it by $r^{13/4}$;
one gets: $r^{-13/4}\ig_{\D_r}(\ig|p'|)^{5/4}\le
C(r\r^{-1})^{1/2}K_\r$, which yields the first term of the second
inequality in \equ(4.4). Complete the derivation of the second of
\equ(4.4). (\Idea Note that $p''(\xx,t)$ can be written, in the
interior of $B_r$, as $p''=\tilde p+\hat p$ with:
$$\tilde p(\xx)=-\fra1{4\p}\ig_{B_\r}\fra{\V x-\V y}{|\V x-\V y|^3}\f(\V
y)\W  u\cdot\W \dpr\,\V u\,d\V y,\qquad
\hat p(\xx)=-\fra1{4\p}\ig_{B_\r}\fra{\V\dpr\f(\V y)\cdot(\W u\cdot\W\dpr)\V
u}{|\xx-\yy|}\,d\V y$$
(always at fixed $t$ and not declaring explicitly the
$t$--dependence). Hence by using $|\xx-\yy|>\r/4$, for $\xx\in
B_r$ and $\yy\in B_\r/B_{3\r/4}$, \ie for $\yy$ in the part  of
$B_\r$ where $\V\dpr\,\f\ne\V0$) we find
$$\eqalign{
\ig_{B_r}|\tilde p|\,d\V x\le&\, C\ig_{B_\r}d\V
y\Big(\ig_{B_r}\fra{d\V x}{|\V x-\V y|^2}\,|\V u(\V y)|\,
|\V\dpr\,\W u(\V y)|\Big)\le
\cr\le&\,C r \Big(\ig_{B_\r}|\V u|^2\Big)^{1/2}\Big(\ig_{B_\r}
|\W\dpr\V u|^2\Big)^{1/2}\le
C r \r^{1/2}A_\r^{1/2}\Big(\ig_{B_\r}|\W\dpr\V u|^2\Big)^{1/2}\cr
\ig_{B_r}|\hat p|\,d\V x\le&\, C\fra{r^3}{\r^2}\ig_{B_\r}
|\V u||\W\dpr\V u|\le C\,
r\r^{1/2} A_\r^{1/2}\Big(\ig_{B_\r}|\W\dpr\V u|^2\Big)^{1/2}\cr}$$
and $\Big(\ig_{B_r}|p''|\Big)^{5/4}$ is bounded by raising the
right hand sides of the last inequalities to the power $5/4$ and
integrating over $t$, and finally applying inequality (H) to generate
the integral $\Big(\ig_{Q_\r}|\W\dpr\V u|^2\Big)^{5/8}$).

\*
\0{\bf[4]:} Deduce that \equ(4.5) holds for a pseudo--NS field
$(\V u,p)$,
\cfr definition 1. (\Idea Let $\f(\V x,t)$ be a $C^\io$ function
which is $1$ on $Q_{\r/2}$ and $0$ outside $Q_{\r}$; it is: $0\le
\f(\V x,t)\le1$, $|\V\dpr\f|\le \fra{C}\r$,
$|\D\f|+|\dpr_t\f\le\fra{C}{\r^2}$, if we suppose that $\f$ has the
form $\f(\xx,t)=\f_2(\fra{\xx}{\r},\fra{t}{\r^2})\ge0$ for some $\f_2$
suitably fixed and smooth. Then, by applying the third of \equ(3.1)
and using the notations of the preceding problems, if $\bar t\in
\D_{\r/2}(t_0)$, it is
$$\eqalignno{
&\ig_{B_r\times\{\bar t\}}|\V u(\xx,t)|^2d\xx\le
\fra{C}{\r^2}\ig_{Q_\r}dtd\xx\,|\V u|^2+\ig_{Q_\r}dtd\xx\,(|\V
u|^2+2p)\,\V u\cdot\V\dpr\f \le\cr
&\le\fra{C}{\r^2}\ig_{Q_\r}dtd\xx\,|\V u|^2+\Big|\ig_{Q_\r}dtd\xx\,
(|\V u|^2-\lis{|\V u|^2_\r})\V u\cdot\V\dpr\f\Big|+
2\ig_{Q_\r} dtd\xx\,p\,\V u\cdot\V\dpr\f\le\cr
&\le\fra{C}{\r^{1/3}}\Big(\ig_{Q_\r}dtd\xx\,|\V u|^3\Big)^{2/3}
+\Big|\ig_{Q_\r}dtd\xx\,\Big(|\V u|^2-
\lis{|\V u|^2_\r}\Big)\,\V u\cdot\V\dpr\f\Big|+\fra{2C}\r\ig_{B_\r}
dtd\xx\,|p||\V u|\le\cr &\le C\r G_\r^{2/3}+C\r J_\r+
\r\Big|\fra1\r\ig_{Q_\r}dtd\xx\,(|\V u|^2-\lis{|\V u|^2_\r})\,
\V u\cdot\V\dpr\f\Big|&(*)\cr}$$
We now use the following inequality, at $t$ fixed and with the
integrals over $d\xx$
$$\eqalign{
&\fra1\r\Big|\ig_{B_\r}\,d\xx\,(|\V u|^2-\lis{|\V u|^2_\r}) \V
u\cdot\V\dpr\f\Big|\le
\fra{C}{\r^2}\ig_{B_\r} \,d\xx\,|\V u|\,\Big||\V
u|^2-\lis{|\V u|^2_\r} \Big|\le\cr
&\le\fra{C}{\r^2}\Big(\ig_{B_\r}\,d\xx\,|\V
u|^3\Big)^{1/3}\Big(\ig_{B_\r}|\V u^2-\lis{|\V u|^2_\r}|^{3/2}
\Big)^{2/3}\cr}$$
%
and we also take into account inequality (P) with $f=\V u^2$ and
$\a=3/2$ which yields (always at $t$ fixed and with integrals over
$d\xx$):
$$\Big(\ig_{B_\r}\Big|\V u^2-\lis{|\V u|^2_\r}\Big|^{3/2}
\Big)^{2/3}\le
C\,\Big(\ig_{B_\r}|\V u||\W\dpr\V u|\Big)$$
then we see that
$$\eqalign{ &\ig_{B_\r}\Big||\V u|^2-\lis{|\V u|^2_\r}\Big|\,|\V
u|\,|\W\dpr\V \f|\le\fra{C}\r\Big(\ig_{B_\r}|\V
u|^3\Big)^{1/3}\Big(\ig_{B_\r}|\V u||\W\dpr\V u|\Big)\le\cr&
\le\fra{C}\r\Big(\ig_{B_\r}|\V u|^3\Big)^{1/3}\Big(\ig_{B_\r}|\V
u|^2\Big)^{1/2}\Big(\ig_{B_\r}|\W\dpr\V u|^2\Big)^{1/2}
\le\cr&\le\fra{C}\r\r^{1/2}A_\r^{1/2}
\Big(\ig_{B_\r}\V u|^3\Big)^{1/3}\cdot
\Big(\ig_{B_\r}|\W\dpr\V u|^2\Big)^{1/2}\,\cdot1\cr}$$
Integrating over $t$ and applying  (H) with exponents $3,2,6$,
respectively, on the last three
factors of the right hand side we get
$$\fra1{\r^2}\ig_{Q_\r}|\V u|\Big|\,|\V u|^2-\lis{|\V
u|^2_\r}\Big|\le C A_\r^{1/2}G_\r^{1/3}\d_\r^{1/2}\le
C\,(G_\r^{2/3}+A_\r\d_\r)$$
and placing this in the first of the preceding inequalities (*) we
obtain the desired result).
\*

{\it The following problems provide a guide to the proof of theorem
II. Below we replace, unless explicitly stated the sets $B_r,Q_r,\D_r$
introduced in definition 2, in (C) above, and employed in the previous
problems with $B^0_r,\D^0_r,Q^0_r$ with $B^0_r=\{\V x|\,|\V x-\V x_0|<
r\}$, $\D^0_r=\{t\,|\, t_0>t>t_0-r^2\}$, $Q^0_r=\{(\V x,t)|\,|\V x-\V
x_0|< r,\, t_0>t>t_0-r^2\}=B^0_r\times\D^0_r$. Likewise we shall set
$B^0_{r_n}=B^0_n,\D^0_{r_n}=\D^0_n. Q^0_{r_n}=Q^0_n$ and we shall define new
operators $A,\d,G,J,K,S$ by the same expressions in
\equ(3.2)\%\equ(3.8) in (C) above but with the  just defined
new meaning of the integration domains. However, to avoid
confusion, we shall call them $A^0,\d^0,\ldots$ with a superscript $0$
added.}
\*

\0{\bf[5]:} With the above conventions check the following
inequalities
$$A^0_n\le C A^0_{n+1},\qquad {G^0_n}\le C {G^0}_{n+1},\qquad {G^0_n}
\le C \,({A^0_n}^{3/2}+{A^0_n}^{3/4}{\d^0_n}^{3/4})$$
(\Idea The first two are trivial consequences of the fact that the
integration domains of the right hand sides are larger than those of
the left hand sides, and the radii of the balls differ only by a
factor $2$ so that $C$ can be chosen $2$ in the first inequality and
$4$ in the second. The third inequality follows from (S) with
$a=\fra34$, $q=3$:
$$\eqalign{
&\ig_{B^0_r}|\V u|^3\le C\,\Big[
\Big(\ig_{B^0_r}|\W\dpr\V u|^2\Big)^{3/4}
\Big(\ig_{B^0_r}|\V
u|^2\Big)^{3/4}+ r^{-3/2}\Big(\ig_{B^0_r}|\V u|^2\Big)^{3/2}\Big]\le
\cr
&\le
C\,\Big[r^{3/4}
{A^0_r}^{3/4}\Big(\ig_{B^0_r}|\W\dpr\V u|^2\Big)^{3/4}+
{A^0_r}^{3/2}\Big]\cr}$$
where the integrals are over $d\xx$ at $t$ fixed; and integrating over
$t$ we estimate $G^0_r$ by applying (H) to the last integral over $t$.)
\*

\0{\bf[6]:} Let $n_0=n+p$ and $Q^0_n=\{(\V x,t)|\,|\V x-\V x_0|<
r_n,\, t_0>t>t-r_n^2\}{\buildrel def\over =} B^0_n\times\D^0_n$ consider
the function:
$$\f_n(\xx,t)=\fra{\exp(-(\xx-\xx_0)^2/4(r_n^2+t_0-t))}{(4\p
(r_n^2+t_0-t))^{3/2}},\qquad(\xx,t)\in Q^0_{n_0}$$
and a function $\chi_{n_0}(\xx,t)=1$ on $Q^0_{n_0-1}$ and $0$ outside
$Q^0_{n_0}$, for instance choosing, a function which has the form
$\chi_{n_0}(\xx,t)=\tilde f(r_{n_0}^{-1}\xx,r_{n_0}^{-1/2}t)\ge0$,
with $\widetilde\f$ a $C^\io$ function fixed once and for all.  Then
write \equ(3.1) using $\f=\f_n\chi_{n_0}$ and deduce the inequality
$$\fra{{A^0_n}+{\d^0_n}}{r_n^2}\le
C\Big[r_{n+p}^{-2}{G^0_{n+p}}^{2/3}+\sum_{k=n+1}^{n+p}r_k^{-2}
{G^0_k}+r_{n+p}^{-2}{J^0}_{n+p}+
\sum_{k=n+1}^{n+p-1} r_k^{-2}L_k\Big]\eqno(@)$$
where $L_k=r_k^{-2}\ig_{Q^0_k}d\xx\,dt\,|\V u||p-\lis{p^k}|$ with
$\lis{p^k}$ equals the average of $p$ on the ball ${B^0_k}$; for each
$p>0$. (\Idea Consider the function $\f$ and note that $\f\ge(C
r_n^3)^{-1}$ in $Q^0_n$, which allows us to estimate {\it from below}
the left hand side term in \equ(3.1), with $(C
r_n^2)^{-1}({A^0_n}+{\d^0}_n)$. Moreover one checks that
$$\eqalign{
&|\f|\le \fra{C}{r_m^3},\qquad
|\V\dpr\f|\le\fra{C}{r_m^4},\quad n\le m\le n+p\=n_0,
\quad{\rm in}\quad Q^0_{m+1}/Q^0_m\cr
&|\dpr_t\f+\D\f|\le\fra{C}{r_{n_0}^5}\kern4.5cm{\rm in}\quad Q^0_{n_0}\cr}$$
and the second relation follows from $\dpr_t\f+\D\f\=0$ in the
``dangerous places'', \ie $\chi=1$, because $\f$ is a solution of the
heat equation (backward in time). Hence the first term in the right hand
side of \equ(3.1) can be bounded {\it from above} by
$$\ig_{Q^0_{n_0}}|\V
u|^2|\dpr_t\f_n+\D\f_n|\le\fra{C}{r_{n_0}^5}\ig_{Q^0_{n_0}}|\V
u|^2\le\fra{C}{r_{n_0}^5} \Big(\ig_{Q^0_{n_0}}|\V
u|^3\Big)^{2/3}r_{n_0}^{5/3}\le\fra{C}{r_{n_0}^2}{G^0}_{n_0}^{2/3}$$
getting the first term in the r.h.s. of (@).

Using here the scaling properties of the function $\f$ the second term
is bounded by
$$\eqalign{
&\ig_{Q^0_{n_0}}|\V u|^3|\Dpr\f_n|\le\fra{C}{r_n^4}\ig_{Q^0_{n+1}}|\V
u|^3+\sum_{k=n+2}^{n_0}\fra{C}{r_k^4}\ig_{Q^0_k/Q^0_{k-1}}|\V u|^3\le
\cr
&\le\sum_{k=n+1}^{n_0}\fra{C}{r_k^4}\ig_{Q^0_k}|\V u|^3\le
C\sum_{k=n+1}^{n_0}\fra{G^0_k}{r_k^2}\cr}$$
Calling the third term (\cfr \equ(1.3)) $Z$ we see that it is bounded
by
$$\eqalign{
Z&\le \Big|\ig_{Q^0_{n_0}} p\,\V
u\cdot\Dpr\,\chi_{n_0}\f_n\Big|\le
\Big| \ig_{Q^0_{n+1}} p\,\V u\cdot\Dpr\,\chi_{n+1}\f_n\Big|
+\cr
&+\sum_{k=n+2}^{n_0}
\Big| \ig_{Q^0_{k}} p\,\V u\cdot\Dpr\,(\chi_k-\chi_{k-1})\f_n\Big|\le
\Big| \ig_{Q^0_{n+1}} (p-\lis{p^{n+1}})\,\V u\cdot\Dpr\,\chi_{n+1}\f_n\Big|
+\cr
&+\sum_{k=n+2}^{n_0-1}
\Big| \ig_{Q^0_{k}} (p-\lis{p^k})
\,\V u\cdot\Dpr\,(\chi_k-\chi_{k-1})\f_n\Big|
+\ig_{Q^0_{n_0}}|\V u|\,|p|\,|\Dpr(\chi_{n_0}-\chi_{n_0-1})\f_n\cr}$$
where $\lis{p^m}$ denotes the average of $p$ over $B^0_m$ (which only
depends on $t$): the possibility of replacing $p$ by $p-\lis p$ in the
integrals is simply due to the fact that the $0$ divergence of $\V u$
allows us to add to $p$ {\it any} constant because, by integration by
parts, it will contribute $0$ to the value of the integral.

{}From the last inequality it follows
$$Z\le \sum_{k=n+1}^{n_0-1}\fra{C}{r_k^4}\ig_{Q^0_k}|p-\lis{p^k}|\,|\V u|
+{J^0_{n_0}} r_{n_0}^{-2}=
\sum_{k=n+1}^{n_0-1}\fra{C}{r_k^2}L_k\,+{J^0_{n_0}} r_{n_0}^{-2}$$
then sum the above estimates.)
\*

\0{\bf[7]} If $\V x_0$ is the center of $\O$ the function $\chi_{n_0}p$
can be regarded, if $n_0<-1$, as defined on the whole $R^3$ and zero
outside the torus $\O$. Then if $\D$ is the Laplace operator {\it on
the whole} $R^3$ note that the expression of $p$ in terms of $\V u$
(\cfr (a) of \equ(3.1)) implies that in $Q^0_{n_0}$:
$$\chi_{n_0}p=\D^{-1}\D\chi_{n_0}p\=\D^{-1}\Big(p\D\chi_{n_0}+
2(\Dpr\chi_{n_0})\cdot(\Dpr p)-\chi_{n_0}\W\dpr\V\dpr\cdot(\W u\V u)
\Big)$$
Show that this expression can be rewritten, for
$n<n_0$, as
$$\eqalign{
\chi_{n_0}p=&-\D^{-1}(\ch_{n_0} \V\dpr\W\dpr(\V u\W u))+
[\D^{-1}(p\D \ch_{n_0})+2 (\V\dpr\D^{-1})((\V\dpr
\ch_{n_0}) p)]=\cr
&=[-(\V\dpr\W\dpr\D^{-1})(\ch_{n_0}\V u\W u)]+
[2(\V\dpr\D^{-1})(\W\dpr\ch_{n_0}\V u\W
u)-\D^{-1}((\V\dpr\W\dpr\ch_{n_0})\V u\W u)]+\cr&+
[\D^{-1}(p\D \ch_{n_0})+2 (\V\dpr\D^{-1})((\V\dpr
\ch_{n_0}) p)]\defi p_1+p_2+p_3+p_4\cr}$$
with $p_1=-(\W\dpr\V\dpr\D^{-1})(\chi_{n_0}\th_{n+1}\V u\W u)$ and
$p_2=-(\W\dpr\V\dpr\D^{-1})(\chi_{n_0}(1-\th_{n+1})\V u\W u)$ where
$\th_k$ is the characteristic function of ${B^0}_k$ and $p_3,p_4$ are
the last two terms in square brackets. (\Idea Use, for $\xx,t\in
Q^0_{n_0}$, Poisson formula
$$\eqalign{
\chi_{n_0}(\xx,t) p(\xx,t)=&\fra{-1}{4\p}\ig_{B^0_{n_0}}
\fra{\D\,((\chi_{n_0}p)(\yy,t))}{|\xx-\yy|}d\yy=\cr
&=\fra{-1}{4\p}\ig_{B^0_{n_0}}\fra{
p\D\chi_{n_0}+
2\Dpr\chi_{n_0}\cdot\Dpr p-\chi_{n_0}\W\dpr\V\dpr\cdot(\W u\V u)
}{|\xx-\yy|}d\yy\cr}$$
and suitably integrate by parts).
\*

\0{\bf[8]} In the context of the previous problem
check that the formulae derived there can be written more explicitly
as
$$\eqalign{
p_1=&-(\W\dpr\Dpr\D^{-1})\cdot\big(\chi_{n_0}\th_{n+1}\V u\,\W
u\big),\quad p_2=-\fra1{4\p}\ig_{B^0_{n_0}/{B^0}_{n+1}}
\big(\Dpr\W\dpr\fra1{|\xx-\yy|}\big)\cdot\chi_{n_0}\V u\,\W u
\cr
p_3=&\fra1{2\p}\ig_{B^0_{n_0}}
\fra{\xx-\yy}{|\xx-\yy|^3}(\W\dpr\chi_{n_0})\V u \,\W
u+\fra1{4\p}\ig_{B^0_{n_0}}
\fra1{|\xx-\yy|}(\W\dpr\Dpr\chi_{n_0})\W u\,\V u
\cr
p_4=&-\fra{1}{4\p}\ig\fra1{|\xx-\yy|}p(\yy)\D\chi_{n_0}
+\fra{2}{4\p}\ig p(\yy)\fra{\xx-\yy}{|\xx-\yy|^3}\cdot\V\dpr\chi_{n_0}\cr}$$
where $n< n_0$ and the integrals are over $\yy$ at $t$ fixed, and the
functions in the left hand side are evaluated in $\xx,t$.
\*

\0{\bf[9]:} Consider the quantity $L_n$, introduced in [6],
$$L_n\defi r^{-2}_n\ig_{Q_n}|\V u|\ |p-\lis p_n(\th)|\,d\V\x d\th$$
and show that, setting  $n_0=n+p,\,p>0$, it is
$$\eqalign{ L_n&\le C\Big[\Big({r_{n+1}\over
r_{n_0}}\Big)^{7/5}{A^0_{n+1}}^{1/5}{G^0_{n+1}}^{1/5}{K^0_{n_0}}^{4/5}
+\Big({r_{n+1}\over
r_{n_0}}\Big)^{5/3}{G^0_{n+1}}^{1/3}{G^0_{n_0}}^{2/3}+\cr
&+{G^0_{n+1}}+r^3_{n+1}{G^0_{n+1}}^{1/3}
\sum^{n_0-1}_{k=n+1}r^{-3}_k{A^0_k}
\Big]\cr}$$
(\Idea Refer to [8] to bound $L_n$ by: $\sum_{i=1}^4
r_n^{-2}\ig_{Q_n^0}|\V u||p_i-\lis{p_i^n}|$ where $\lis{p^n_i}$ is the
average of $p_i$ over $B^0_n$; and estimate separately the four
terms. For the first it is not necessary to subtract the average and
the difference $|p_1-\lis{p^{n}_1}|$ can be divided into the sum of
the absolute values each of which contributes equally to the final
estimate which is obtained via the (CZ), and the (H)
$$\ig_{B^0_{n+1}}|p_1-\lis{p_1}||\V u|\le 2
\Big(\ig_{{B^0_{n+1}}}|p_1|^{3/2}\Big)^{2/3}\Big(\ig_{B^0_{n+1}}|\V
u|^3\Big)^{1/3}\le C\ig_{B^0_{n+1}}|\V u|^3$$
and the contribution of $p_1$ at $L_n$ is bounded, therefore, by $C
G^0_{n+1}$: note that this would not be true with $p$ instead of $p_1$
because in the right hand side there would be $\ig_\O|p|^{3/2}$ rather
than $\ig_{B^0_{n+1}}|p|^{3/2}$, because the (CZ) is a ``nonlocal''
inequality.  The term with $p_2$ is bounded as
$$\eqalign{
&\ig_{{\D^0_n}^0}\ig_{B^0_n}|p_2-\lis{p^n_2}||\V
u|\le\ig_{{\D^0_n}^0}\ig_{B^0_n}|\V u|\,r_n\max|\Dpr p_2|\le\cr
&\le r_n\Big(\ig_{Q^0_n}\fra{|\V u|^3}{r_n^2}\Big)^{1/3}
r_n^{2/3}r_n^{10/3}\max_{Q^0_n}|\Dpr p_2|\le\cr
&\le r_n^5 {G^0_n}^{1/3}\sum_{m=n+1}^{n_0-1}\max_{t\in\D^0_m}
\ig_{B^0_{m+1}/B^0_m}\fra{|\V u|^2}{r_m^4} = r_n^5
{G^0_n}^{1/3}\sum_{m=n+1}^{n_0-1}\fra{A^0_m}{r_m^3}\cr}$$
Analogously the term with $p_3$ is bounded by using $|\Dpr p_3|\le C
r_{n_0}^{-4}\ig_{B^0_{n_0}}|\V u|^2$ which is majorized by $C
r_{n_0}^{-3}$ $(\ig_{B^0_{n_0}}|\V u|^3)^{2/3}$ obtaining
$$\eqalign{
&\fra1{r_n^2}\ig_{Q^0_n}|\V u||p_3-\lis{p^n_3}|\le\fra{C}{r_n^2}
r_{n_0}^{-3}\ig_{\D^0_n}
[(\ig_{B^0_{n}}|\V u|^3)^{2/3} r_n\ig_{B^0_n}|\V u|]\le\cr
&
\le\fra{C}{r_n^2} r_n^3r_{n_0}^{-3}
\ig_{\D^0_n}(\ig_{B^0_{n}}|\V
u|^3)^{2/3}(\ig_{B^0_{n}}|\V u|^3)^{1/3}\le\cr
&\le \fra{C}{r_n^2}(\fra{r_n}{r_{n_0}})^3 r_{n_0}^{4/3}r_n^{2/3}
{G^0_{n_0}}^{2/3}{G^0_n}^{1/3}=C
(\fra{r_n}{r_{n_0}})^{5/3}{G^0_{n_0}}^{2/3}{G^0_n}^{1/3}\cr}$$
Finally the term with $p_4$ is bounded (taking into account that the
derivatives $\D\ch_n,\Dpr\ch_n$ vanish where the kernels become bigger
than what suggested by their dimension) by noting that
$$\ig_{B^0_n}|p_4-\lis{p_4^n}||\V u|\le C r_n\ig_{B^0_n}
|\V u| \max_{B^0_n}|\Dpr
p_4|\le C r_n\Big(\ig_{B^0_n}|\V u|\Big)\Big(\ig_{B^0_{n_0}}
\fra{|p|}{r_{n_0}^4}\Big)$$
Denoting with ${\tilde K}^0_{n_0}$ the operator $K^0_{n_0}$ without the
factor $r_{n_0}^{-13/4}$ which makes it dimensionless, and
introducing, similarly, ${{\tilde
A}^0_{n}},{\tilde G}^0_n$ we obtain the
following chain of inequalities, using repeatedly (H)
$$\eqalign{
&\fra1{r_n^2}\ig_{Q^0_n}|p_4-\lis{p_4^n}||\V u|\le
\fra{C}{r_n^2}r_n\Big(\ig_{\D^0_n}
\Big(\ig_{B^0_{n_0}}\fra{|p|}{r_{n_0}^4}\Big)^{5/4}
\Big)^{4/5}
\Big(\ig_{\D^0_n}\Big(\ig_{B^0_n}|\V
u|\Big)^5\Big)^{1/5}\le\cr &\le\fra{C}{r_n^2}\fra{r_n}{r_{n_0}^4}
 {{\hbox{$\displaystyle\tilde K$}^0_{n_0}}}^{\kern-4pt4/5}
\Big(\ig_{\D^0_n}\Big(\ig_{B^0_n}|\V
u|^{2/5}|\V u|^{3/5}\cdot1\Big)5\Big)^{1/5}\le\cr &
\le\fra{C}{r_n^2}\fra{r_n}{r_{n_0}^4}
{{\hbox{$\displaystyle\tilde K$}^0_{n_0}}}^{\kern-4pt4/5}
\Big(\ig_{B^0_n}|\V u|^2\Big)^{1/5}\Big(\ig_{Q^0_{n}}|\V
u|^3\Big)^{1/5} r_n^{9/5}\le\cr
&\le \fra{C}{r_n^2}\fra{r_n}{r_{n_0}^4}r_n^{12/5}
{\hbox{$\displaystyle\tilde K$}^0_{n_0}}^{\kern-4pt4/5}
{\hbox{$\displaystyle \tilde A$}^0_n}^{1/5}
{\hbox{$\displaystyle \tilde G$}^0_n}^{1/5}
\le C\Big(\fra{r_n}{r_{n_0}}
\Big)^{7/5} {A^0_n}^{1/5}{G^0_n}^{1/5}{K^0_{n_0}}^{\kern-4pt4/5}\cr}$$
Finally use the inequalities of [5] and combine the estimates
above on the terms $p_j,\,j=1,..,4$.)
\*

\0{\bf[10]} Let $T_n=(A^0_n+\d^0_n)$; combine inequalities
of [6] and [9], and [5] to deduce
$$\eqalignno{
T_n\le&\,2^{2n}\Big(2^{-2n_0}\e+\sum^{n_0-1}_{k=n+1}2^{-2k}T^{3/2}_k+
2^{-2n_0}\e+2^{-7{n_0}/5}\e\sum^{n_0-1}_{k=n+1}2^{-3k/5}T^{1/2}_k+
\Big.\cr
&\Big.+\e2^{-5{n_0}/3}\sum^{n_0-1}_{k=n_0+2}2^{-k/3}T^{1/2}_k+
\sum^{n_0-1}_{k=n+1}2^{-2k}T^{3/2}_k+
\sum^{n_0-1}_{k=n+1}2^{k}T^{1/2}_k\sum^{n_0-1}_{p=k}2^{-3p}
T_p\Big)\cr
\e&\, \equiv C\max({G^0_{n_0}}^{\kern-4pt2/3},
{K^0_{n_0}}^{\kern-4pt4/5},{J^0}_{n_0})\cr}$$
and show that, by induction, if $\e$ is small enough then
$r_n^{-2}T_n\le \e^{2/3}r_{n_0}^{-2}$.
\*

\0{\bf[11]:} If $G(r_0)+J(r_0)+K(r_0)<\e_s$ with $\e_s$ small
enough, then given $(\xx',t')\in {Q}_{r_0/4}(\V x_0,t_0)$, show that
if one calls $G^0_r,J^0_r,K^0_r,A^0_r,\d^0_r$ the operators associated
with $Q^0_r(\V x',t')$ then
$$\limsup_{n\to\io}\fra1{r_n^2} A^0_n\le C \fra{\e_s^{2/3}}{r_0^2}$$
for a suitable constant $C$. (\Idea Note that
$Q^0_{r_0/4}(\xx',t')\subset {Q}_{r_0}(\xx_0,t_0)$ hence
$G^0_{r_0/4},J^0_{r_0/4},\ldots$ are bounded by a constant, ($\le
4^2$), times $G(r_0),J(r_0)$\-$\ldots$ respectively. Then apply the
result of [10]).
\*

\0{\bf[12]:} Check that the result of [11] implies theorem II.
(\Idea Indeed
$$\fra1{r_n^2}A^0_n\ge \fra1{r_n^3}\ig_{B^0_n}|\V
u(\xx,t')|^2d\xx\tende{n\to-\io}
\fra{4\p}3\,|\V u(\xx',t')|^2$$
where $B^0_n$ is the ball centered at $\xx'$, {\it for almost all} the
points $(\xx,',t')\in Q^{0}_{r_0/4}$; hence $|\V u(\xx',t')|$ is
bounded in $Q^0_{r_0/4}$ and one can apply proposition 2).
\*

\0{\bf[13]:} Let $f$ be a function with zero average over
$B^0_r$. Since $f(\xx)=f(\yy)+ \ig_0^1 ds\,\Dpr
f(\yy+(\xx-\yy)s)\cdot(\xx-\yy)$ for each $\yy\in B^0_r$, averaging
this identity over $\yy$ one gets
$$f(\xx)=\ig_{B^0_r}\fra{d\yy}{|B^0_r|}
\,\ig_0^1 ds\,\Dpr f(\yy+(\xx-\yy)s)\cdot(\xx-\yy)$$
Assuming $\a=1$ prove (P). (\Idea Change variables
as $\yy\to\V z=\yy+(\xx-\yy)s$ so that for $\a$ integer
$$\ig_{B^0_r}|f(\xx)|^\a\fra{d\xx}{|B^0_r|}\=
\ig_{B^0_r}\fra{d\xx}{|B^0_r|}\Big|\ig_0^1\ig_{B^0_r}
\fra{d\V z}{|B^0_r|}\,\fra{ds}{(1-s)^3}
\Dpr f(\V z)\cdot(\V z-\xx)\Big|^\a$$
where the integration domain of $\V z$ depends from $\xx$ and $s$, and
it is contained in the ball with radius $2(1-s)r$ around $\xx$. The
integral can then be bounded by
$$\ig\fra{d \V z_1}{|B^0_r|}\fra{d s_1}{1-s_1}\ldots
\fra{d \V z_\a}{|B^0_r|}\fra{d s_\a}{1-s_\a} (2r)^\a |\Dpr f(\V
z_1)|\ldots|\Dpr f(\V z_\a)|\,\ig \fra{d\xx}{|B^0_r|}$$
where $\xx$ varies in a domain with $|\xx-\V z_i|\le 2 (1-s_i)r$ for
each $i$. Hence the integral over $\fra{d\xx}{|B^0_r|}$ is bounded by
$8(1-s_i)^3$ for each $i$. Performing a geometric average of such
bounds (over $\a$ terms)
$$\eqalign{
&\ig_{B^0_r}|f(\xx)|^\a\fra{d\xx}{|B^0_r|}\le 2^{\a+3} r^\a
\prod_{i=1}^\a\ig\fra{d\V z_i\,ds_i}{|B^0_r|(1-s_i)}||\Dpr f(\V z_i)|\,
(1-s_i)^{3/\a}\le\cr
&\le 2^{\a+3} r^\a\Big(\ig_{B^0_r}|\Dpr f(\V z)|\fra{d\V z}{|B^0_r|}\Big)^\a
\cdot\Big(\ig_0^1\fra{ds}{(1-s)^{3-3/\a}}\Big)^\a\cr}$$
getting (P) for $\a=1$ and an explicit estimate of the constant
$C^P_1$: this also gives a heuristic motivation for (P) with $\a<\fra32$.)
\*

\0{\bf[14]:} Differentiate twice with respect to $\a^{-1}$ and check
the convexity of $\a^{-1}\to||f||_{\a}\= (\ig
|f(\xx)|^\a\,d\xx/|B^0_r|)^{1/\a}$. Use this to get (P) for each
$1\le\a<\a_0$ if it is valid for $\a=\a_0$.  
(\Idea Since (P) can be written $||f||_\a\le
C_\a\,(\ig|\Dpr f|\,d\xx/r^2)$ then if $\a^{-1}=\th
\a_0^{-1}+(1-\th)(\a_0+1)^{-1}$ with $\a_0$ integer it follows that
$C_\a$ can be taken $C_\a=\th C_{\a_0}+(1-\th)\,C_{\a_0+1}$).
\*

\0{\bf[15]:} Consider a sequence $\V u^\l$ of solutions of the Leray
regularized equations which converges {\it weakly} (\ie for each
Fourier component) to a Leray solution.  By construction the $\V
u^\l,\V u$ verify the \ap bounds in
\equ(1.2) and (hence) \equ(2.5).  Deduce that $\V u$ verifies the
\equ(3.1). (\Idea Only (c) has to be proved. Note that if
$\V u^\l\to\V u^0$ weakly, then the left hand side of \equ(1.3) is
semi continuous hence the value computed with $\V u^0$ is not larger
than the limit of the right hand side in \equ(1.3).  On the other
hand the right hand side of \equ(1.3) is {\it continuous} in the
limit $\l\to\io$. Indeed given $N>0$ weak convergence implies
$$\eqalign{
\lim_{\l\to\io }&\ig_0^{T_0}dt\ig_\O |\V u^\l-\V u^0|^2\,d\V
x\=\lim_{\l\to\io }
\ig_0^{T_0} dt \sum_{0<|\V k|}|\V\g_\kk^\l(t)-\V\g^0_\kk(t)|^2\le\cr
\le\lim_{\l\to\io }&\big(\sum_{0<|\kk|<N}\ig_0^{T_0} dt
|\V\g_\kk^\l(t)-\V\g^0_\kk(t)|^2+\sum_{|\kk|\ge N}\ig_0^{T_0} dt
{|\kk|^2\over N^2}|\V\g_\kk^\l(t)-\V\g^0_\kk(t)|^2\big)\le\cr
\le\lim_{\l\to\io }&\big(\sum_{0<|\kk|<N}\ig_0^{T_0} dt
|\V\g_\kk^\l(t)-\V\g^0_\kk(t)|^2+{1\over N^2}\ig_0^{T_0} dt \ig_\O
|\W\dpr(\V u^\l-\V u^0)|^2\big)=\cr
=\lim_{\l\to\io }&{1\over N^2}\ig_0^{T_0} dt \ig_\O
|\W\dpr(\V u^\l-\V u^0)|^2\le {2E_0\n^{-1}\over N^2}\cr}$$
using the \ap bound in \equ(1.2) (with zero force) and componentwise
convergence of the Fourier transform $\V\g_\kk(t)$ of $\V u(t)$
to the Fourier transform $\g^0_\kk(t)$ of $\V u^0$. Hence
$\ig_0^{T_0}\ig_\O|\V u^\l-\V u^0|^2\to0$ showing the convergence of
the first two terms of the right hand side of \equ(1.3) to the
corresponding terms of (c) in \equ(3.1).

Apply, next, the inequality (S), \equ(2.2), with
$q=3,\,a={3\over4},\,{q\over 2}-a={3\over4}$, and again by the \ap
bounds in \equ(1.2) we get
$$\eqalign{
&\ig_0^{T_0} dt\ig_\O|\V u^\l-\V u^0|^3 \,d\V x\le C\ig_0^{T_0}
dt\,||\W\dpr(\V u^\l-\V u^0)||^{3/2}_2\,||\V u^\l-\V u^0||^{3/2}_2\le\cr
&\le C\Big(\ig_0^{T_0}dt\, ||\W\dpr(\V u^\l-\V u^0)||^2_2\Big)^{3/4}
\,\Big(\ig_0^{T_0} dt\,||\V u^\l-\V u^0||^6_2\Big)^{1/4}\le\cr &\le
C(2E_0\n^{-1})^{3/4}(2\sqrt{E_0})\ig_0^{T_0} dt\,||\V u^\l-\V
u^0||_2^2\tende{\l\to\io }0\cr}$$
showing continuity of the third term in the second member of
\equ(3.1).  Finally, and analogously, if we recall that
$p^\l=-\D^{-1}\sum_{ij}\dpr_i\dpr_j(u_i^\l u_j^\l)$ and if we apply
the inequalities (CZ) and (H), we get
$$\eqalign{
&\txt\ig_0^{T_0} dt\ig_\O d\V x|p^\l\V u^\l-p^0\V u^0|\le\ig\ig
|p^\l-p^0|\,|\V u^\l|+\ig\ig  |p^0|\,|\V u^\l-\V u^0|\le\cr
&\txt\Big(\ig\ig |p^\l-p^0|^{3/2}\Big)^{2/3}\,\Big(\ig\ig|\V
u^\l|^3\Big)^{1/3}+
\Big(\ig\ig |p^0|^{3/2}\Big)^{2/3}\,\Big(\ig\ig|\V u^\l-\V
u^0|^3\Big)^{1/3}\cr}$$
where the last integral tends to zero
by the previous relation while
the first, via (CZ), will be such that
$\ig_0^{T_0}\ig_\O|p^\l-p^0|^{3/2}\le\Big(\ig\ig|\V u^\l-\V
u^0|^3\Big)^{2/3}\tende{\l\to\io }0$
proving the continuity of the fourth term in the right hand side of (c)
in \equ(3.1). Hence the right hand side is continuous in the
considered limit).
\*

\0{\bf[16]: } ({\it covering theorem}, (Vitali)) Let $S$ be an
arbitrary set inside a sphere of $R^n$. Consider a {\it covering} of
$S$ with little open balls with the {\it Vitali property}: \ie such
that every point of $S$ is contained in a family of open balls of the
covering whose radii have a zero greatest lower bound. Given $\h>0$
show that if $\l>1$ is large enough it is possible to find a
denumerable family $F_1,F_2, \ldots$ of pairwise disjoint balls of the
covering with diameter $<\h$ such that $\cup_i \l F_i\supset S$ where
$\l F_i$ denotes the ball with the same center of $F_i$ and radius $\l$
times longer. Furthermore $\l$ can be chosen independent of $S$, see
also problem [17]. (\Idea Let $\FF$ be the covering and let $a=max_\FF
\,diam(F)$.  Define $a_k=a 2^{-k}$ and let $\FF_1$ be a {\it maximal}
family of {\it pairwise disjoint} ball of $\FF$ with radii $\ge
a2^{-1}$ and $<a$. Likewise let $\FF_2$ be a maximal set of balls of
$\FF$ with radii between $a 2^{-2}$ and $a 2^{-1}$ pairwise disjoint
between themselves and with the ones of the family
$\FF_1$. Inductively we define $\FF_1,\ldots,\FF_k,\ldots$. It is now
important to note that if $x\not\in\cup_k \FF_k$ it must be:
$distance(x,\FF_k)<\l a 2^{-k}$ for some $k$, if $\l$ is large
enough. If indeed $\d$ is the radius of a ball $S_\d$ containing $x$
and if $a 2^{-k_0}\le \d<a 2^{-k_0+1}$ then the point of $S_\d$
farthest away from $x$ is at most at distance $\le 2\d<4 a 2^{-k_0}$;
and if, therefore, it was $d(x,\FF_{k_0})\ge 4 a 2^{-k_0}$ we would
find that the set $\FF_{k_0}$ could be made larger by adding to it
$S_\d$, against the maximality supposed for $\FF_{k_0}$. Note that
$\l=5$ is a possible choice.)
\*

\0{\bf[17]: } Show that if the balls in problem [16] are replaced
by the {\it parabolic cylinders} which are Cartesian products of a
radius $r$ ball in the first $k$ coordinates and one of radius
$r^\a$, with $\a\ge1$ in the $n-k$ remaining ones, then the result
still holds if one replaces $5F_i$ with $\l F_i$ where $\l$ is a
suitable homothety factor (with respect to the center of $F_i$). Show
that if $\a=1,2$ then $\l=5$ is enough (and, in general,
$\l=(4^2+2^{2(1+\a)/\a})^{1/2}$ is enough).
\*

\0{\bf[18]: } Check that the Hausdorff dimension of the Cantor set $C$
is $\log_3 2$, \cfr \equ(5.3). (\Idea It remains to see, given the
equation in footnote${}^{9}$, that if $\a<\a_0$ then
$\m_\a(C)=\io$. If $\d=3^{-n}$ the covering $\CC_n$ of $C$ with the
$n$--th generation intervals is ``the best'' among those with sets of
diameter $\le 3^{-n}$ because another covering could be refined by
deleting from each if its intervals the points that are out of the
$n$--th generation intervals.  Furthermore the inequality $1<2
3^{-\a}$ for $\a<\log_32$ shows that it will not be convenient to
further subdivide the intervals of $\CC_n$ for the purpose of
diminishing the sum $\sum |F_i|^\a$. Hence for $\d=3^{-n}$ the minimum
value of the sum is $2^n 3^{-n\a}\tende{n\to\io}\io$.)
\vfil}
\*

\0{\bf Bibliography:}
\*\*

\0{[BG95] } {\bf Benfatto, G., Gallavotti, G.}: {\sl  Renormalization
group}, p. 1--144, Princeton University Press, 1995.
\*
\0{[CKN82] } {\bf Caffarelli, L., Kohn, R., Nirenberg, L
}: {\it Partial regularity of suitable weak solutions of the Navier-
Stokes equations}, Communications on pure and applied mathematics, 35,
771- 831, 1982.
\*
\0{[CF88a] } {\bf Constantin, P., Foias, C.}: {\sl  Navier
Stokes Equations}, Chicago Lectures in Mathematics series, University
of Chicago Press, 1988.
\*
\0{[DS60] } {\bf Dunford, N., Schwartz, I.}: {\sl Linear
operators}, Interscience, 1960.
\*

\0{[Ga02] } {\bf Gallavotti, G.}: {\sl Foundations of Fluid
Mechanics}, p. 1--513, Springer-Verlag, Berlin, 2003.
\*

\0[LL01] {\bf Lieb, E., Loss, M.}: {\sl Analysis}, American
Mathematical Society, second edition, Providence, 2001.
\*

\0{[PS87] } {\bf Pumir, A., Siggia, E.}: {\it Vortex dynamics
and the existence of solutions to the Navier Stokes equations},
Physics of Fluids, {\bf30}, 1606--1626, 1987.
\*
\0{[Sc77] } {\bf Scheffer, V.}: {\it Hausdorff dimension
and the Navier Stokes equations}, Communications in Mathematical
Physics, 55, 97- 112, 1977.  And {\it Boundary regularity for the
Navier Sokes equation in half space}, Communications in Mathematical
Physics, {\bf 85}, 275- 299, 1982.
\*
\0{[So63] } {\bf Sobolev, S.L.}: {\sl Applications of Functional
analysis in Mathematical Phy\-sics}, Translations of the American
Mathematical Society, vol 7, 1963, Providence.
\*
\0{[St93] } {\bf Stein, E.}: {\sl Harmonic analysis}, Princeton
University press, 1993.
\bye